\documentclass[12pt]{article}

\usepackage[inline,shortlabels]{enumitem}
\usepackage{float}
\usepackage{mathtools}
\usepackage{graphicx}
\usepackage[round]{natbib}
\usepackage{setspace}

\usepackage[top=1in, left=1in, right=1in, lines=42]{geometry}
\usepackage{tikz}
\usepackage[english]{babel}
\usepackage{longtable}
\usepackage{color}
\usepackage{amssymb,amsmath,amsthm}
\usepackage{multirow}
\usepackage[titletoc,title]{appendix}
\usepackage{authblk}
\usepackage{setspace}
\usepackage{dsfont}
\usepackage[OT1]{fontenc}
\usepackage{refcount}
\usepackage{booktabs}
\graphicspath{ {./figs/} }

\usepackage{nameref,zref-xr} 
\zxrsetup{toltxlabel} 

\usepackage{subfig}

\newtheorem{theorem}{Theorem}
\AtEndDocument{\refstepcounter{theorem}\label{finalthm}}
\AtEndDocument{\refstepcounter{equation}\label{finaleq}}

{
  \theoremstyle{definition}
  
}
{
  \theoremstyle{definition}
  \newtheorem{assumptioniden}{}
}
{
  \theoremstyle{definition}
  
}

\newtheorem{lemma}{Lemma}
\AtEndDocument{\refstepcounter{lemma}\label{finallemma}}

\newtheorem{definition}{Definition}

 \DeclareMathOperator{\var}{\mathsf{Var}}

\renewcommand{\P}{\mathsf{P}}
\newcommand{\F}{\mathsf{F}}

\newcommand{\p}{\mathsf{p}}

\newcommand{\indep}{\mbox{$\perp\!\!\!\perp$}} 
 \newcommand{\dd}{\mathrm{d}}

\newcommand{\one}{\mathds{1}}
 
\newcommand{\E}{\mathsf{E}}

\renewcommand{\P}{\mathsf{P}}
\newcommand{\D}{\mathsf{D}}
\newcommand{\T}{\mathsf{T}}
\newcommand{\U}{\mathsf{U}}

\newcommand{\CATE}{f}

\renewenvironment{proof}{{\it Proof }}{\qed \\}

\DeclarePairedDelimiterX{\norm}[1]{\lVert}{\rVert}{#1}

\newtheorem{proposition}{Proposition}

\pgfdeclarelayer{background}
\pgfsetlayers{background,main}
\usetikzlibrary{arrows,positioning}
\tikzset{
>=stealth',
punkt/.style={
rectangle,
rounded corners,
draw=black, very thick,
text width=6.5em,
minimum height=2em,
text centered},
pil/.style={
->,
thick,
shorten <=2pt,
shorten >=2pt,}
}
\newcommand{\Vertex}[2]
{\node[minimum width=0.6cm,inner sep=0.05cm] (#2) at (#1) {$\footnotesize#2$};
}
\newcommand{\Vertexr}[2]
{\node[rectangle, draw, minimum width=0.6cm,inner sep=0.05cm] (#2) at (#1) {$\footnotesize#2$};
}
\newcommand{\ArrowR}[3]%
{ \begin{pgfonlayer}{background}
\draw[->,#3] (#1) to[bend right=30] (#2);
\end{pgfonlayer}
}
\newcommand{\ArrowL}[3]%
{ \begin{pgfonlayer}{background}
\draw[->,#3] (#1) to[bend left=45] (#2);
\end{pgfonlayer}
}
\newcommand{\EdgeL}[3]%
{ \begin{pgfonlayer}{background}
\draw[dashed,#3] (#1) to[bend right=-45] (#2);
\end{pgfonlayer}
}

\newcommand{\Arrow}[3]%
{ \begin{pgfonlayer}{background}
\draw[->,#3] (#1) -- +(#2);
\end{pgfonlayer}
}

\newcommand{\DHArrow}[3]%
{ \begin{pgfonlayer}{background}
\draw[<->,#3,dashed] (#1)to[bend right=30](#2);
\end{pgfonlayer}
}
\setlist[itemize]{noitemsep} 
\usepackage{enumitem}
\usepackage{url}
\title{Improving efficiency in transporting average treatment effects}

\date{}
\author[1]{Kara E. Rudolph\thanks{corresponding author:}}
\author[1]{Nicholas Williams}
\author[2]{Elizabeth Stuart}
\author[3]{Iv\'an D\'iaz}
\affil[1]{\small Department of
  Epidemiology, Mailman School of Public Health, Columbia University.}
\affil[2]{\small Department of Biostatistics, Johns Hopkins University.}
\affil[3]{\small Division of Biostatistics, New York University Grossman School of Medicine.}

\newcommand{\B}{\mathsf{B}}
\usepackage{array}
\begin{document}
\vspace{-2cm}
\singlespacing{
\maketitle

\begin{abstract}
We develop flexible, semiparametric estimators of the average treatment effect (ATE) transported to a new population (``target population'') that offer potential efficiency gains.  Transport may be of value when the ATE may differ across populations. We consider the setting where differences in the ATE are due to differences in the distribution of baseline covariates that modify the treatment effect (``effect modifiers''). First, we propose a collaborative one-step semiparametric estimator that can improve efficiency. This approach does not require researchers to have knowledge about which covariates are effect modifiers and which differ in distribution between the populations, but does require all covariates to be measured in the target population. Second, we propose two one-step semiparametric estimators that assume knowledge of which covariates are effect modifiers and which are both effect modifiers and differentially distributed between the populations. These estimators can be used even when not all covariates are observed in the target population; one requires that only effect modifiers are observed, and the other requires that only those modifiers that are also differentially distributed are observed.  
We use simulation to compare finite sample performance across our proposed estimators and an existing semiparametric estimator of the transported ATE, including in the presence of practical violations of the positivity assumption. Lastly, we apply our proposed estimators to a large-scale housing trial. 
\end{abstract}

\noindent%
{\it Keywords:}transportability; efficiency; generalizability; effect modification; causal inference
}
\newpage
\section{Introduction}
\label{sec:intro}
\doublespacing
The effect of a treatment or exposure on an outcome may differ from one sample or population to another. Such treatment effect heterogeneity drives concerns about external validity---specifically, in the presence of such treatment effect heterogeneity, a treatment effect estimated in a study sample may not apply to a target population. 
In this paper, we focus on the 
setting where we have observed data $O=(S, W, A, S \times Y),$ where $S$ is an indicator of the population and represents all factors that produce differences between the two populations \citep{pearl2015generalizing}; $S=1$ denotes the source population and $S=0$ denotes the target population; $W$ denotes a vector of baseline covariates; $A$ denotes treatment/ exposure; 
and $Y$ denotes the outcome, only observed in the source population. We consider data structures that satisfy $S-$ignorability (and $S-$admissibility), meaning that differences in the effect of $A$ on $Y$ across populations in $S$ are due to differences in the distribution of baseline covariates that also modify the treatment effect. In this setting, the transported average treatment effect (transported ATE) is defined as the average effect of $A$ on $Y$ transported to the target population, denoted $\E(Y_1 - Y_0 \mid S=0),$ where $Y_{a}$ denotes the counterfactual outcome had treatment been set to value $a$, possibly counter to fact. Others have shown that the transported ATE is identified by the statistical parameter $\E[\E(Y \mid A=1, W, S=1) - \E(Y \mid A=0, W, S=1) \mid S=0]$ under the following assumptions. First, the so-called ``transport'' assumption or $S-$admissibility: $\E(Y \mid A, W, S=1) = \E(Y \mid A, W,S=0)$ \citep{pearl2011transportability}. Second, the conditional exchangeability assumption: $A  \indep Y_a \mid W, S=0$. And third, positivity. 
We note that this identification result also holds under the alternative $S-$admissibility assumption that $\E(Y_a \mid W, S=1) = \E(Y_a \mid W, S=0)$, and the alternative conditional exchangeability assumption that $A  \indep Y_a \mid W, S=1$. We discuss when one would use one set of assumptions vs. the other in Section \ref{sec:note}.

There are many estimators of the transported ATE, including those based on inverse probability weighting \cite[e.g.,][]{cole2010generalizing,li2018balancing}, matching \cite[e.g.,][]{stuart2011use,bennett2020building}, g-computation \cite[e.g.,][]{dahabreh2019generalizing}, and doubly robust semiparametric estimators based on the efficient influence function \cite[e.g.,][]{rudolph2017robust}. \citet{rudolph2017robust} derived the efficient influence function (EIF) 
for estimation of transported effects; inspection of the EIF shows that the semiparametric efficiency bound is inversely proportional to the probability of being in the source population conditional on covariates. This means that, in designs with limited overlap between the distribution of baseline covariates across populations, standard semiparametric efficient estimators may suffer from large variability (one can see evidence of this lack of precision in the transported ATE in our motivating example in Figure \ref{mtoresfig}). This motivates the need to develop transport estimators 
with improved precision. 

In this paper, we show that the precision of transported estimates may be improved by considering two types of covariate dimension reduction; we develop an estimator for each approach. 
First, we propose a so-called ``collaborative'' estimator that 
reduces the dimension of the transport problem to improve efficiency. We call this estimator ``collaborative'' following the insights of \citet{benkeser2020nonparametric,van2010collaborative}, who exploit the fact that the propensity score in a doubly robust estimator only needs to adjust for the outcome variability not explained by the outcome model in order to achieve double robustness. 
In our problem, this means that the probability of being in the source population only needs to be adjusted by a transformation of the covariates given by the conditional average treatment effect (CATE). We show that an estimator constructed using this insight provides efficiency gains compared to standard semiparametric estimators. This estimator is agnostic to the causal structure in the sense that it does not require knowledge of the variables that influence the CATE, it only requires a consistent estimator of the CATE. 
Second, we construct two other semiparametric estimators for cases where the causal structure underlying effect heterogeneity is known. These estimators make use of the fact that 
it is not necessary to use all covariates in the transport step; it is also not necessary to use all the effect modifiers in the transport step. Only effect modifiers which actually differ across populations are needed. 

The estimators we propose contribute to previous and concurrent work that also sought to improve the efficiency of transported estimates \citep{egami2021covariate,zeng2023efficient}. 
\citet{egami2021covariate} considered a scenario where the source population consists of unconditionally randomized data and proposed an estimator that couples parametric causal learning with inverse probability weighting, requiring the full covariate vector, $W$, to be present in both populations. \citet{zeng2023efficient} assume knowledge of which subset of baseline covariates 
are needed for transport and propose a semiparametric estimator for transported ATE under the model that incorporates that knowledge. Their proposal is a special case of our second approach, which we detail further in Section \ref{sec:estknownv}

This paper is organized as follows. First, we give notation and the standard identification assumptions 
in Section 2. In Section 3, we propose a collaborative semiparametric estimator that can improve efficiency but does not require information on which variables are effect modifiers.  
Then, in Section 4, we consider the case where the investigator has knowledge of which variables are effect modifiers 
and potentially also which of these 
differ in distribution across $S$. We show an identification result incorporating this knowledge in Section 4.1. In Section 4.2,  
we derive a doubly robust estimating equation 
from which we develop estimators that can use machine learning for estimation of the nuisance parameters. In Section 4.3, we propose 
a semiparametric estimator 
based on the estimating equation, and in Section 4.4, using the asymptotic variance of this estimator, we outline scenarios where incorporating this knowledge 
results in semiparametric efficiency gains and scenarios where incorporating these assumptions results in efficiency losses.  
 In Section 5, we illustrate finite sample performance of our proposed estimators in several scenarios using simulation. In Section 6, we apply our proposed estimators to data from the Moving to Opportunity Study (MTO). 
Section 7 concludes. 

\vspace{-.5cm}
\section{Preliminaries}
\label{sec:note}
We assume the same observed data as discussed in the Introduction: $O=(S, W, A, S \times Y)$, and assume $O_1, \ldots, O_n$ represents a sample of $n$
i.i.d.~observations of $O$. We assume the following nonparametric structural equation model: 
    $S=f_S(U_S); W=f_W(S, U_W); A=f_A(S,W,U_A); 
    Y=f_Y(W, A, U_Y),$
where each variable is a deterministic, unknown function, $f$, of unobserved exogenous errors, $(U_S, U_W, U_A, U_S)$, as well as possibly observed endogenous variables \citep{Pearl2009}. We assume $S$ can be separated into a source population, $S=1$, and a target population, $S=0$. We also assume $A$ denotes a binary treatment/exposure variable. $Y$ can be binary or numerical. 

We use $\P$ to
denote the distribution of $O$. $\P$ is an element of the nonparametric statistical
model defined as all continuous densities on $O$. 
We let $\E$ denote
expectation, and define
$\P f = \int f(o)\dd \P(o)$ for a given function $f(o)$.  

As stated in the Introduction, without further structural causal assumptions, and under positivity and the following conditional exchangeability and $S-$admissibility assumptions:
\begin{assumptioniden}[Conditional exchangeability]\label{ass:conf}
  $Y_a \indep A\mid W, S=0,$ 
\end{assumptioniden}
\begin{assumptioniden}[$S-$admissibility / Transportability of the outcome model]\label{ass:exch}
  $\E(Y\mid A, W, S=1) = \E(Y\mid A, W,S=0)$,
\end{assumptioniden}
\noindent it is well known that the transported ATE of a binary treatment can be identified as 
$$ \E(Y_1 - Y_0 \mid S=0) = \E[\E(Y \mid A=1, w, S=1) - \E(Y \mid A=0, w, S=1) \mid S=0].$$ We provide the proof in Section S1 of the Supplement. We also note that one could alternatively identify the transported ATE of a binary treatment under alternative conditional exchangeability and $S-$admissibility assumptions:

\noindent \textbf{Alternative A1} (Conditional exchangeability) $Y_a \indep A\mid W, S=1$, \\
\noindent \textbf{Alternative A2} ($S-$admissibility). $\E(Y_a \mid W,S=1) = \E(Y_a \mid W, S=0)$. 

For example, if $S=1$ represents trial data where the exchangeability assumption is likely to hold and $S=0$ represents observational data where the exchangeability assumption is less likely to hold, then using these alternative versions would be preferable. However, in the motivating example we consider here where $S=1$ and $S=0$ both represent trial data, we prefer the assumptions as written above, as this allows us to write $S-$admissibility as a function of observed data.

For convenience in some of the results and methods we propose, we will parameterize the outcome conditional expectation as 
\[\E(Y\mid A, W, S=1)=A\times f(W) + g(W),\] 
where, under the identification assumptions listed above, $f(W)$ represents the conditional average treatment effect (CATE) in the source population, $\E(Y\mid A=1, W, S=1) - \E(Y\mid A=0, W, S=1) = f(W)$, and $g(W)$ represents the conditional outcome  expectation among the untreated in the source population, $\E(Y\mid A=0, W, S=1)$. This parameterization was also used by \citet{chernozhukov2018double,nie2021quasi,hahn2020bayesian,colnet2021generalizing}. Using this notation, we can write the statistical estimand as
$$\lambda = \E\{f(W)\mid S=0\}.$$

\vspace{-.5cm}

\section{Collaborative estimator for the transported ATE}
\label{sec:estunknownv}
We first propose a collaborative semiparametric estimator that 
leverages two main insights for covariate dimension reduction to improve efficiency. First, we rely on the fundamental role of the propensity score to find a dimension-reduction of the covariates that is sufficient for transporting to the target population \citep{rosenbaum1983central}. Second, we rely on the insight that the probability of being in the source population only needs to be adjusted by the CATE 
in order to transport to the target population. We first provide intuition for each of these two underpinnings of the proposed estimator, and defer a more rigorous presentation of the proposed estimator to Section \ref{sec:estima}.

\subsection{The fundamental role of the propensity score}
Let $e_S(w)=\P(S=1\mid W=w)$ denote the ``propensity'' to be in study $S=1$ (what we also refer to as the source population) conditional on covariates $W$. Note that because 
$S\indep W\mid e_S(W)$ \citep{rosenbaum1983central}, where $e_S(W)$ denotes the propensity score $\P(S=1 \mid W)$, we can 
reparameterize  $\lambda$ as 
\[\lambda = \E[f(W)\mid S=0]=\E\{\E[f(W)\mid e_S(W), S=0]\mid S=0\} = \E\{\E[f(W)\mid e_S(W)]\mid S=0\}.\] Intuitively, this reparameterization shows that the propensity score $e_S(W)$ is an appropriate data-reduction of the variables $W$ that is sufficient to transport the ATE. That is, once the CATE $f(W)$ is consistently estimated, it suffices to transport its relation to 
this propensity score. We will show in Section \ref{sec:estima} how this can be used to obtain efficiency gains. 

\subsection{The fundamental role of the conditional average treatment effect}
Using the tower rule, we can obtain the following result:
\[\lambda = \E[f(W)\mid S=0]=\E\left[\frac{\one\{S=1\}}{\P(S=0)}\frac{(1-h_S(W))}{h_S(W)}f(W)\right],\]
where we define $h_S(W)=\P(S=1 \mid f(W))$. Thus, an using inverse probability weighed estimator would only need to re-weight observations using the probability $h_S(W)$ for transport instead of $e_S(W)$. Because the inverse weights based on $h_S(W)$ condition on a univariate transformation of the covariates, namely the CATE, they are expected to be more stable than the inverse weights based on $e_S(W)$, which condition on the full covariate vector $W$, leading to efficiency gains.

\vspace{-.5cm}
\subsection{Proposed estimator}\label{sec:estima}

To summarize, our proposed estimator 
can be thought of as ``collaborative", because instead of estimating the propensity of being in the source population 
conditional on $W$, $\P(S=1 \mid W)$, 
it estimates this propensity as a function of treatment effect heterogeneity (i.e., collaboratively, based on how well it ``tunes'' an estimate of the CATE in the source population \citep{benkeser2020nonparametric}): 
$\P(S=1 \mid f(W))$. In addition, instead of estimating the CATE in the source population, 
$f(W)$, it projects an estimate of the CATE onto the propensity score for being in the source population, 
$\E\{f(W) \mid e_S(W)\}$ (i.e., collaboratively, based on how well it ``tunes'' an estimate of the propensity score). 

Our estimation strategy relies on obtaining an approximation of the first-order bias of a plug-in estimator of $\lambda$, also known as a von Mises expansion \citep{mises1947asymptotic} and defined as follows:
\begin{definition}[First-order von Mises expansion]\label{def:vonmises}
    A function $\U(O;\P)$ that depends on the data and a distribution $\P$ satisfies a first-order von Mises expansion if the following holds $\lambda(\F) -\lambda(\P) = -\E_\P[\U(O;\F)] + R(\F,\P)$
    for any two distributions $\F$ and $\P$, where $R(\F,\P)$ is a second-order term given by sums of terms taking the form $\int c(\F,\P)\{a(\F)-a(\P)\}\{b(\F)-b(\P)\}\dd\P$ for transformations $a$, $b$, and $c$.
\end{definition}
\label{sec:eeunknownv}

Consider an estimate $\hat \P$ of the distribution of the data obtained, for example, with machine learning. Then, according to Definition \ref{def:vonmises}, the bias of the plug-in estimator $\lambda(\hat \P)$ for estimating $\lambda(\P)$ is given by $-\E_\P[\U(O;\hat\P)]+R(\hat \P, \P)$. If the second-order term $R(\hat \P, \P)$ can be made small, then the bias of the plug-in estimator is dominated by the first-order term $-\E_\P[\U(O;\hat\P)]$, which can be estimated through an empirical mean $-n^{-1}\sum_i\U(O_i;\hat\P)$. This idea underlies the construction of the estimators proposed in this manuscript. 


We introduce some additional notation to simplify the presentation of the results. Let $e_A(w)=\P(A=1\mid W=w, S=1)$ and $\p = \P(S=0)$.
We define $k(w)=\E\{f(W)\mid e_S(W)=e_S(w)\}$, and let 
  \begin{align*}
	\U_{\lambda}(O;\P) &= \frac{\one\{S=1\}}{\p}\left\{\frac{A}{e_A(W)}-\frac{1-A}{1-e_A(W)}\right\}\frac{1-h_S(W)}{h_S(W)}\{Y -
                    	A f(W) - g(W)\}\\
                    	&+\frac{\{1-e_S(W)\}}{\p}[f(W) - k(W)]\\
    	&+\frac{\one\{S=0\}}{\p}[k(W)-\lambda],
  \end{align*}
where $\lambda$ was defined in Section \ref{sec:note}. Because $\U_\lambda(\cdot;\P)$ depends on $\P$ through $\eta=(e_A, e_S, h_S, f, k, g,\p)$, we will sometimes use the notation $\U_\lambda(\cdot;\eta)$. 
We note that to distinguish between the non-collaborative and collaborative estimators for $\lambda$, we denote the collaborative estimator for $\lambda$ estimated using $\U_\lambda$ as $\tilde{\lambda}_C$.

\begin{theorem}[First-order bias of plug-in estimator] \label{theo:alt}
For any estimator $\hat\eta=(\hat e_A, \hat e_S, \hat h_S, \hat f, \hat k, \hat g)$, 
let $\hat\lambda_C$ denote the plug-in collaborative estimator defined as 
\[\hat\lambda_C = \frac{1}{n\times \hat \p}\sum_{i=1}^n\one\{S_i=0\}\hat k(W_i).\]
The bias of $\hat\lambda_C$ is equal to
$\hat\lambda_C - \lambda = -\E[\U_{\lambda}(O;\hat \eta)] + R_U(\eta, \hat \eta),$
where $R_U$ is a second-order term (defined in the proof of the theorem) equal to a sum of products of errors of the type $\int c(\eta,\hat\eta)(a(\hat\eta) - a(\eta))(b(\hat\eta) - b(\eta))\dd\P$. The proof is included in Section S2 of the Supplement.
\end{theorem}

\vspace{-.5cm}
\subsection{Estimation}
\label{sec:collabest}
We propose a collaborative one-step semiparametric estimator of $\lambda$ that solves the $U_{\lambda}(O,\hat\eta)$ estimating equation. 
One-step estimators are constructed by subtracting the first-order bias estimate from the plug-in estimator: $\tilde{\lambda}_C(O,\hat\eta) = \hat{\lambda}_C(O, \hat \P) + \frac{1}{n}\sum^n_{i=1}\U_{\lambda}(O_i, \hat \eta)$. 
This estimator relies on consistent estimation of $f(W)$ (so is not doubly robust). 

In what follows, we make use of the following doubly robust unbiased transformation to estimate the CATE \citep{rubin2007doubly,luedtke2016super}. 
\begin{lemma}[Doubly robust unbiased transformation for estimation of the CATE, $f(W)$] 
\label{lemma:dr}For any
  distribution $\P_1$, define
  \begin{align*}
    \T(O;\P_1) &= \frac{2A-1}{\P_1(A\mid S=1,
      W)}\{Y - \E_1(Y\mid A, W, S=1)\}\\ & + \E_1(Y\mid A=1, W, S=1)-
    \E_1(Y\mid A=0, W, S=1).
  \end{align*}
  If $\P_1$ is such that $\P_1(A\mid S=1,
  W)=\P(A\mid S=1,
  W)$ or $\E_1(Y\mid A, W, S=1)=\E(Y\mid A, W, S=1)$, we have that $\E\{\T(O;\P_1)\mid W, S=1\}=f(W)$. \end{lemma}


This estimator can be implemented as follows. 
\begin{enumerate}[topsep=0pt,itemsep=-1ex,partopsep=1ex,parsep=1ex]
\item First, estimate nuisance parameters $\hat e_A(W) = \hat\P(A\mid S=1, W)$ and
  $\hat\E(Y\mid A, W, S=1)$. One could use parametric regression models (e.g., generalized linear models) or data-adaptive regressions that incorporate machine learning in model fitting for these regressions and for those that follow below. We do the latter in the simulations, motivating example, and in the software we provide. 
\item Use the nuisance parameter estimates to calculate $\T(O_i;\hat\P)$ among observations with $S_i =1$.
\item Regress $\T(O_i;\hat\P)$ on $W$ among observations with $S_i=1$. The resulting predicted values 
are estimates of $f(W_i)$, denoted $\hat f(W_i)$. 

\item Regress $S$ on $\hat f(W)$. The resulting predicted values are estimates of $h_S(w)$, denoted $\hat h_S(w)$. 
\item Estimate the nuisance parameter $\hat e_S(W) = \hat \P(S=1 \mid W)$ using regression. 


\item Regress $\hat f(W)$ on $\hat e_S(W)$. The predicted values, $\hat k(W)=\hat\E(\hat f(W) \mid \hat e_S(W))$, reflect study heterogeneity in the CATEs.
\item Compute an initial estimator $\hat \lambda_C = \frac{1}{\sum_{i=1}^n
    \one\{S_i=0\}}\sum_{i=1}^n
  \one\{S_i=0\}\hat k(W_i)$.
\item Compute estimates of $g(W_i)$ for all observations $i$, denoted $\hat g(W_i)$, by generating predicted values from the estimator of $\E(Y\mid A=0, W, S=1)$.  

\item Compute the one-step estimator as $\tilde\lambda_C = \hat \lambda_C
  +\frac{1}{n}\sum \U_{\lambda}(O_i;\hat \eta)$, where $\hat \eta$ is comprised of $\hat e_A$, $\hat e_S$, $\hat f$, $\hat g$, $\hat h_S$, $\hat k$, and $\hat\p$.
  \item Last, we can estimate the variance of $\tilde\lambda_C$ as the sample variance of $\U_{\lambda}(O_i;\hat \eta)$.
\end{enumerate}

We use a cross-fitted version of the above estimator to avoid relying on the Donsker class assumption to achieve asymptotic normality \citep{klaassen1987consistent,zheng2011cross, chernozhukov2016double}. 
Let ${\cal V}_1, \ldots, {\cal V}_J$
denote a random partition of data with indices $i \in \{1, \ldots, n\}$ into $J$
prediction sets of approximately the same size such that 
$\bigcup_{j=1}^J {\cal V}_j = \{1, \ldots, n\}$. For each $j$,
the training sample is given by
${\cal T}_j = \{1, \ldots, n\} \setminus {\cal V}_j$. 
$\hat \eta_{j}$ denotes the estimator of $\eta$, obtained by training
the corresponding prediction algorithm using only data in the sample
${\cal T}_j$, and $j(i)$ denotes the index of the
validation set which contains observation $i$. Then use these fits, $\hat\eta_{j(i)}(O_i)$ in computing  $\U_\lambda(O_i, \hat\eta_{j(i)})$.  

This estimator is asymptotically normal, licensing the construction of Wald-type confidence intervals and hypothesis tests under the assumptions of the following theorem.

\begin{theorem}[Weak convergence of $\tilde\lambda_C$]
    Assume the second-order term $R_U(\eta,\hat \eta)=o_P(n^{-1/2})$, and that $\hat e_S(W)$ and $\hat e_A(W)$ are bounded away from zero in probability. Then we have $\sqrt{n}(\tilde \lambda_C - \lambda)\rightsquigarrow N(0;\xi^2),$
    where $\xi^2=\var[\U_{\lambda}(O;\hat \eta)]$.\end{theorem}

The proof of this theorem is an application of Theorem~\ref{theo:alt} above and Proposition 1 of \citet{kennedy2022semiparametric}. Because $R_U$ is a second-order term, the assumption $R_U(\eta,\hat \eta)=o_P(n^{-1/2})$ is expected to hold, for instance, if each component of $\eta$ is estimated at a consistency rate of $n^{1/4}$ or faster. This is significantly slower that the parametric convergence rate $n^{1/2}$, and allows us to use flexible regression estimators from the statistical learning literature such as regression trees, boosting, parametric models using $\ell_1$ penalization, etc.

Software to implement this estimator is available for download from 
\url{https://github.com/nt-williams/transport} (function \texttt{transport\_ate\_incomplete\_sans\_Z()}). 

This approach to collaborative estimation, which builds on 
\citet{benkeser2020nonparametric}'s construction of a collaborative estimator of the non-transported ATE, is simpler than alternative approaches to collaborative estimation \citep[e.g.,][]{van2010collaborative,schnitzer2018collaborative}. But, that simplicity comes at a price. Although our proposed estimator is asymptotically linear, it is not doubly robust---we rely on consistent estimation of $f(W)$. In addition, like the collaborative estimator of \citet{benkeser2020nonparametric}, it is super efficient---meaning that its asymptotic variance is less than the efficiency bound---and an irregular estimator, the practical implications of which are poorly understood and the subject of future work. 

\vspace{-.5cm}

\section{Semiparametric RAL estimator incorporating knowledge of the causal structure underlying effect heterogeneity} 
\label{sec:estknownv}

The collaborative estimator of Section \ref{sec:estunknownv} may improve efficiency in estimating the transported ATE without relying on investigator knowledge of which variables influence the CATE. However, if the investigator has knowledge about which covariates are effect modifiers and the further subset that differ in distribution across populations, then we show in this Section how this knowledge can be used to 1) achieve identification of the transported ATE in cases where the full vector $W$ is not measured in the target population and 2) offer semiparametric efficiency gains with respect to the previous semiparametric estimator \citep{rudolph2017robust} that does not use this knowledge 
(including finite sample efficiency gains due to a weaker positivity assumption, given below, than that required for identification of $\lambda$). 

\begin{figure}[H]
       \caption{Directed acyclic graph (DAG) where: $A$ represents treatment, $Y$ represents outcome, $S$ represents population, $W$ represents baseline covariates and can be further categorized into i) effect modifiers on the additive scale, $V \subseteq W$; ii) covariates that differ in distribution across $S$, $X \subseteq W$; and iii) covariates that are both effect modifiers \textit{and} differ in distribution across $S$, 
$Z = V \cap X$. This DAG depicts the 
differential distribution of effect modifiers, $Z$, across $S$, and differential distribution of non-effect modifiers, $X\backslash Z$, across $S$. 
$Z$ is a sufficient subset of effect modifiers required for transport.}
     \label{fig:dag}
     \centering
    \includegraphics[width=.5\textwidth]{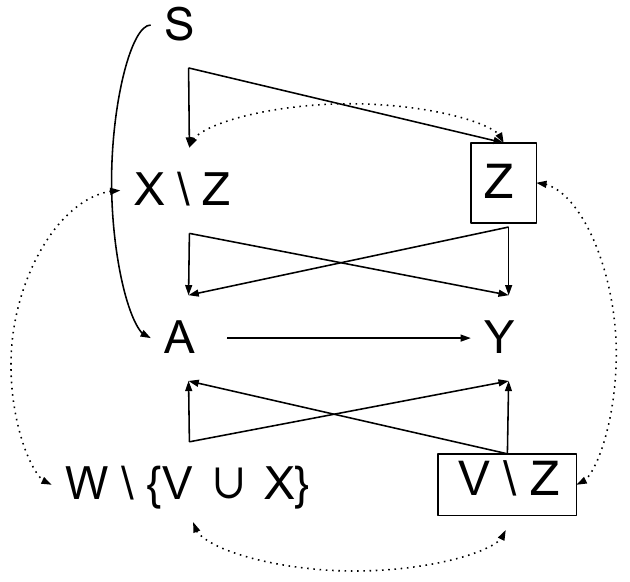}

     \hfill
   \end{figure}

The DAG in Figure \ref{fig:dag} depicts these different subsets of covariates. 
$V$ is the subset of covariates $W$ that are effect modifiers on the additive scale, $V \subseteq W$; $X$ is the subset of $W$ that differ in distribution across $S$, $X \subseteq W$; and $Z$ is the subset that are both effect modifiers \textit{and} differ in distribution across $S$, 
$Z = V \cap X$; thus, $Z\subseteq V \subseteq W$. We follow previous convention and put a box around variable sets to indicate effect modification \citep[e.g.,][]{hernan2002causal}. 
We underscore that this estimation approach requires only a subset of covariates to be measured in the target population. Therefore, it 
be more appealing 
than the estimator in Section \ref{sec:estunknownv} 
when certain variables have not been measured in the target population and it is infeasible to measure them. 

\subsection{Identification}
We make two additional structural assumptions on our statistical model that, together, assume a \textit{sufficient subset of effect modifiers required for transport.}
The first of these two assumptions 
implies that only the variables $V$, where $V\subseteq W$, are modifiers of the ATE, such that $\E(Y_1 - Y_0 \mid W, S=1)  = \E(Y_1 - Y_0 \mid V, S=1)$ under assumptions \ref{ass:conf} and \ref{ass:exch}. Formally, we can state the assumption as:
\begin{assumptioniden}[
Treatment effect modifiers]\label{ass:effmod} The function $f(\cdot)$ depends on $W$ only through 
$V \subseteq W$.
\end{assumptioniden}

The second of these two assumptions implies that only the effect modifiers $Z$, where $Z\subseteq V$, differ in distribution across $S$. Formally, we can state the assumption as: \begin{assumptioniden}[Partial study heterogeneity of effect modifiers]\label{ass:popdiff} Assume there is a subset $Z\subseteq V$ such that $S\indep V\mid Z$.
\end{assumptioniden}

We make a few notes about these additional assumptions. First, assumption \ref{ass:effmod} implies 
\[\E(Y\mid A=1, W, S=1) -\E(Y\mid A=0, W, S=1) = f(V).\] Second, because only the subset $Z$ differs in distribution across $S$, it constitutes a sufficient subset of effect modifiers required for transport. Third, we make assumption \ref{ass:popdiff} with respect to $V$ and not $W$, because the subset of $W$ that differ in distribution between $S=1$ and $S=0$ only matter for transporting if they also modify the treatment effect. Fourth, $Z$ is the intersection of variables that are effect modifiers and variables that differ in distribution across $S$. Fifth, 
 assumptions \ref{ass:effmod} and \ref{ass:popdiff} involve only observed data
distributions; therefore, they are testable. 

For the transported ATE to be well-defined, we assume:
\begin{assumptioniden}[Positivity of study and treatment mechanisms]\label{ass:pos} Assume
  $\P(W=w\mid S=0)>0$ implies $\P(A=a\mid W=w, S=s)>0$ for $s\in\{0,1\}$, which means that there is a positive probability of each value of treatment in the source and target populations conditional on covariates, $w$, that are observed in the target population. 
     $\P(Z=z\mid S=0)>0$ implies $\P(Z=z\mid S=1)>0$, which means each of a sufficient subset of effect modifiers required for transport values, $z$, observed in the target population must also be observed in the source population.
This assumption also depends on only on the observed data, so is also testable.
\end{assumptioniden}


\begin{lemma}\label{lemma:iden} Under assumptions \ref{ass:conf}-\ref{ass:pos}, we have
  $\E(Y_1-Y_0\mid S=0)$ is identified by $\theta$ and $\theta_{\text{alt}}$, where
$\theta = \E[\E\{f(V)\mid Z\}\mid S=0],$\\
\noindent and
$\theta_{\text{alt}} = \E[\E\{f(V)\mid Z, S=1\}\mid S=0].$ We include the proof in Section S1 the Supplement.
\end{lemma}

A particular case of this identification result where $V=Z$ has been shown previously \citep{colnet2021generalizing,zeng2023efficient}. We focus on the statistical parameter $\theta$ for the remainder of the paper, as this allows for pooling data from both the source and target populations in estimating the inner expectation, which is possible in our motivating application since we measured the covariates $V$ in both samples. However, when only $Z$ but not $V\backslash Z$ is measured in the target population, $\theta$ is unidentifiable (we cannot estimate $\E[f(V)\mid Z]$) and $\theta_{\text{alt}}$ must be used instead. We describe an estimator for $\theta_{\text{alt}}$ in Section S3 of the Supplement.

\subsection{Estimating Equation}
\label{sec:eifzknown}

\begin{theorem}
\label{lemma:eif}
 The function $\D_{\theta}(O;\P)$ defined below, satisfies Definition~\ref{def:vonmises} in the
  model with restrictions \ref{ass:effmod} and \ref{ass:popdiff}. 
  \begin{align*}
    \D_{\theta}(O;\P) = \frac{1}{\P(S=0)}\bigg[&\frac{\one\{S=1\}(2A-1)}{\P(A\mid S=1,
                 W)}\frac{\P(S=0\mid Z)}{\P(S=1\mid Z)}\{Y -
                        A f(V) - g(W)\}\\
&+\P(S=0\mid Z)[f(V)-\E\{f(V)\mid Z\}]\\
               &+\one\{S=0\}[\E\{f(V)\mid Z\}-\theta(\P)]\bigg].
  \end{align*}
  \end{theorem}

The function $\D_\theta$ in Theorem \ref{lemma:eif} is similar to the efficient influence function (EIF) for the transported ATE under the more general model that allows all $W$ to modify the additive treatment effect. \citet{rudolph2017robust} show that the nonparametric EIF for $\lambda = \E\{f(W)\mid S=0\}$ is  
equal to the following:
\begin{align*}
  \D_{\lambda}(O;\P) = \frac{1}{\P(S=0)}\bigg[&\frac{\one\{S=1\}(2A-1)}{\P(A\mid S=1,
                       W)}\frac{\P(S=0\mid W)}{\P(S=1\mid W)}\{Y -
                       Af(W) - g(W)\}\\
                     &+\one\{S=0\}\{f(W) -\lambda\} \bigg].
\end{align*}

Comparing $\D_{\theta}$ with $\D_{\lambda}$, we see two main differences: 1) in the first term, $\D_{\theta}$ has weights $\frac{P(S=0 \mid Z)}{P(S=1 \mid Z)}$ whereas $\D_{\lambda}$ has weights $\frac{P(S=0 \mid W)}{P(S=1 \mid W)}$; and 2) $\D_{\theta}$ has an additional term. 

\begin{theorem}[First-order bias of plug-in estimator]\label{theo:lin} 
For any estimator $\hat\gamma=(\hat e_A, \hat e_S, \hat f, \hat g, \hat h, \hat t)$, let $\hat\theta$ denote the plug in estimator defined as 
\[\hat\theta = \frac{1}{n\times \hat \p}\sum_{i=1}^n\one\{S_i=0\}\hat\E[\hat f(V_i) \mid Z_i].\]
The error of $\hat\theta$ is equal to
$\hat\theta - \theta = -\E[\D_{\hat\theta}(O;\hat \gamma)] + R_D(\gamma, \hat \gamma),$
where $R_D$ is a second-order term equal to a sum of products of errors of the type $\int c(\gamma,\hat\gamma)(a(\hat\gamma) - a(\gamma))(b(\hat\gamma) - b(\gamma))\dd\P$.
\end{theorem}
The proof of this theorem is identical to that of Theorem~\ref{theo:alt}, which is included in the Supplement.

\vspace{-.5cm}
\subsection{Estimation}
\label{sec:estimatorknownv}
We propose a one-step semiparametric estimator of $\theta$ under \ref{ass:effmod} and \ref{ass:popdiff}, constructed by subtracting the first-order bias estimate from the plug-in estimator: $\tilde\theta(O, \hat{\gamma}) =  \hat{\theta}(O, \hat \P) + \frac{1}{n}\sum_{i=1}^n \D_\theta(O_i, \hat \gamma).$ 

As in Section \ref{sec:collabest}, we first use a doubly robust unbiased transformation (Lemma \ref{lemma:dr}). Under assumption \ref{ass:effmod}, we have that $\E\{\T(O;\P_1)\mid V, S=1\}=f(V)$, which 
makes our estimator of the CATE doubly robust.

We propose the following one-step estimator: 
\begin{enumerate}[topsep=0pt,itemsep=-1ex,partopsep=1ex,parsep=1ex]
\item First, calculate estimates $\hat e_A(W) = \hat\P(A\mid S=1, W)$ and
  $\hat\E(Y\mid A, W, S=1)$. One could use parametric regression models 
  or data-adaptive regressions that incorporate machine learning in model fitting for these regressions and for those that follow below. We do that latter in the simulations, motivating example, and in the software we provide. 
\item Regress $\T(O_i;\hat\P)$ on $V_i$ among observations with $S_i=1$. The resulting predicted values (generated for all observations $i$) are estimates of $f(V_i)$, denoted $\hat f(V_i)$. 
\item Regress $\hat f(V_i)$ on $Z_i$ in the pooled target and source populations. The resulting predicted values estimate $h(Z)$ and are denoted $\hat h(Z_i)=\hat\E\{\hat f(V_i)\mid Z_i\}$. 
\item Compute an initial estimator $\hat \theta = \frac{1}{\sum_{i=1}^n
    \one\{S_i=0\}}\sum_{i=1}^n
  \one\{S_i=0\}\hat\E\{\hat f(V_i)\mid Z_i\}$.
\item Compute estimates of $g(W_i)$ for all observations $i$, denoted $\hat g(W_i)$, by generating predicted values from the estimator of $\E(Y\mid A=0, W, S=1)$. 
\item Estimate the nuisance parameter $\hat \P(S=1 \mid Z)$ using regression. 
\item Compute the one-step estimator as $\tilde\theta = \hat \theta
  +\frac{1}{n}\sum_i \D_{\hat\theta}(O_i;\hat \gamma)$, where $\hat \gamma$ is comprised of
   $\hat e_A$, $\hat e_S$,
  $\hat f$, $\hat g$, $\hat h$, and $\hat \p$.
  \item Last, we can estimate the variance of $\tilde\theta$ as the sample variance of $\D_{\tilde\theta}(O_i;\hat \gamma)$.
\end{enumerate}


As in the previous section, we use a cross-fitted version of the above estimator. 

We propose use Wald-type estimators for confidence intervals and hypothesis tests, based on the normal distribution and the above variance estimate. These inferential procedures are guaranteed to have correct operating characteristics (e.g., coverage, type I error) under the assumptions of the following theorem.

\begin{theorem}[Weak convergence of $\tilde\theta$]
    Assume the second-order term $R_D(\gamma,\hat \gamma)=o_P(n^{-1/2})$, and that $\hat e_S(W)$ and $\hat e_A(W)$ are bounded away from zero in probability. Then we have
   $\sqrt{n}(\tilde \theta - \theta)\rightsquigarrow N(0;\sigma^2),$
    where $\sigma^2=\var[\D_\theta(O;\gamma)]$.\end{theorem}

 Software to implement this estimator is available for download from \url{https://github.com/nt-williams/transport} (function \texttt{transport\_ate\_incomplete1()}).
 \vspace{-.5cm}
\subsection{Efficiency considerations}\label{section:efficiency}
Under the conditions of Theorem 2 and Result 2 of \citet{rudolph2017robust}, we expect estimators constructed using the estimating equations $\D_\theta$ and $\D_\lambda$ to have asymptotic variance equal to the variance of their respective estimating equations. We therefore now compare the variances $\D_\theta$ with $\D_\lambda$ under assumptions \ref{ass:effmod} and \ref{ass:popdiff} to identify scenarios where we would expect $\theta$ to result in efficiency gains and scenarios where we would expect it to result in efficiency losses. We restrict the comparison to the smaller model under assumptions \ref{ass:effmod} and \ref{ass:popdiff} so that we are comparing the same target parameter (i.e., to make an apples-to-apples comparison). 

\begin{proposition}[Relative efficiency]
  Define
  \[\tau^2(W) = \E\left\{\frac{\P(S=1\mid W)}{\P^2(A\mid W, S=1)}\sigma^2(A, W, S=1)\mid
                W, S=1\right\},\]
 where $\sigma^2(A, W, S)$ is the variance of $Y$ conditional on $(A,
  W, S)$. Then, under assumptions \ref{ass:effmod} and \ref{ass:popdiff}, we have that the variance of $\D_\lambda$ in the
  non-parametric model 
  is equal to
  \begin{align}
  \begin{split}
    &\frac{1}{\p^2}\E\left\{\tau^2(W)\frac{\P^2(S=0\mid
        W)}{\P^2(S=1\mid W)}\right\} + \E\big\{\P(S=0\mid Z)[f(V) - \E\{f(V)\mid Z\}]^2\big\}\\ &+\frac{1}{\p}\E\{[\E(f(V)\mid Z) -\lambda]^2\mid S=0\},\label{eq:effbl}
         \end{split}
  \end{align}
  whereas the variance of $\D_\theta$ in the
  non-parametric model is equal to
  \begin{align}
  \begin{split}
      &\frac{1}{\p^2}\E\left\{\tau^2(W)\frac{\P^2(S=0\mid
        Z)}{\P^2(S=1\mid Z)}\right\} + \E\big\{\P^2(S=0\mid Z)[f(V) - \E\{f(V)\mid Z\}]^2\big\}\\
        &+
    \frac{1}{\p}\E\{[\E(\CATE(V)\mid Z) -\theta]^2\mid S=0\}.\label{eq:effbt}
    \end{split}
    \end{align}
\end{proposition}

Contrasting these two variances reveals in which situations it
might be expected to obtain efficiency gains from using assumptions
\ref{ass:effmod} and \ref{ass:popdiff} in the identification and estimation of the
transported effects. First, note that the last term in both
variances is identical. Consider now the first term in (\ref{eq:effbl}) and (\ref{eq:effbt}). The ratio $\frac{\P(S=0 \mid Z)}{\P(S=1  \mid Z)}$ is expected to be more stable than $\frac{\P(S=0 \mid W)}{\P(S=1  \mid W)}$, so we would expect this to contribute to the $\theta$ estimator having lower variance and better efficiency. However, if $\P(S=1\mid W)$ is
correlated with $\tau^2(W)$, then this correlation may influence efficiency. If the correlation between $\P(S=1\mid Z)$ and
$\tau^2(W)$ is smaller than the correlation between $\P(S=1\mid W)$
and $\tau^2(W)$, then the first term of (\ref{eq:effbt}) might be
larger than the first term in (\ref{eq:effbl}). This could occur, for
example, if the profiles $W$ that have a large conditional variance in
$S=1$ are not well-represented in $S=0$ such that $\P(S=0\mid W)$ is
small. In other words, if $\var(Y \mid A, W,S=1)$ is positively correlated with $\P(S=1 \mid W)$ more than with $\P(S=1 \mid Z)$, then $\P(S=1 \mid W)$ may act to down-weight unstable observations more than $\P(S=1 \mid Z)$ . In this case, we would potentially expect an advantage of the $\lambda$ estimator; whether or not it outweighs the advantage of $\theta$ due to increased stability of $\frac{\P(S=0 \mid Z)}{\P(S=1  \mid Z)}$ depends on specifics of the data-generating mechanism. 
Finally, consider the middle terms in (\ref{eq:effbl}) and (\ref{eq:effbt}). The difference is in $\P(S=0 \mid Z)^2$ for (\ref{eq:effbt}) and $\P(S=0 \mid Z)$ for (\ref{eq:effbl}). The squared term in (\ref{eq:effbt}) will act to decrease the variance and increase the efficiency, adding an efficiency advantage for the $\theta$ estimator regardless of the data-generating mechanism. 
So, to summarize, we would expect the $\theta$ estimator to be more efficient than the $\lambda$ estimator in many cases, especially in the presence of practical positivity violations in finite samples and if the outcome variance is bounded. Practical positivity violations may result if there are certain variables in $W$ that are highly predictive of population membership and consequently, result in near-deterministic predictions of $S=1$ vs. $S=0$. We examine the above scenarios, among others, in simulations in Section \ref{sec:sim}. 

\vspace{-.5cm}

\section{Simulation}\label{sec:sim}

We performed a simulation study to illustrate 
estimator performance in finite sample sizes. We considered four data-generating mechanisms (DGMs), outlined in Table \ref{tab:simdgm}, below. 

\addtolength{\tabcolsep}{2pt} 
\begin{table}[H]
    \centering
    \caption{Data-generating mechanisms considered in the simulation. Note that $W_1$ corresponds to the type of covariate $X \backslash Z$; $W_2$ and $W_3$ correspond to the type of covariate $V \backslash Z$. }
    \footnotesize
    \begin{tabular}{@{}ll@{}}
    \toprule
        DGM 1 & $\begin{aligned}
         \P(W_1 = 1) &= 0.5 \\
         \P(W_2 = 1) &= 0.33 \\
         \P(Z = 1) &= 0.66 \\
         \P(A = 1) &= 0.5 \\
         \P(S = 1 \mid X) &= 0.4 + 0.5W_1 - 0.3Z \\ 
         Y &\sim \mathcal{N}(A + W_1 + AW_2 + 2.5AZ, (0.1 + 0.8W_1)^2)
         \end{aligned}$ \\
    \midrule
        DGM 2 & $\begin{aligned}
         \P(W_1 = 1) &= 0.5 \\
         \P(W_2 = 1) &= 0.33 \\
         \P(Z = 1) &= 0.66 \\
         \P(A = 1) &= 0.5 \\
         \P(S = 1 \mid X) &= 0.5 - 0.4W_1+ 0.3Z \\ 
         Y &\sim \mathcal{N}(A + W_1 + AW_2 + 2.5AZ, (0.1 + 0.5W_1)^2)
         \end{aligned}$ \\
    \midrule
        DGM 3 & $\begin{aligned}
         \P(W_1= 1) &= 0.25 \\
         \P(Z = 1) &= 0.5 \\
         \P(A = 1) &= 0.5 \\
         \P(S = 1 \mid X) &= 0.8-0.6Z-0.18W_1 \\ 
         Y &\sim \mathcal{N}(1.2 + 0.25A + 0.5Z + 0.5W_1 + AZ, 0)
         \end{aligned}$ \\
    \midrule
        DGM 4 & $\begin{aligned}
         \P(W_1 = 1) &= 0.5 \\
         \P(W_2 = 1) &= 0.75 \\
         \P(W_3= 1) &= 0.33 \\
         \P(Z = 1) &= 0.25 \\
         \P(A = 1) &= 0.5 \\
         \P(S = 1 \mid X) &= 0.8 - 0.5Z - 0.25W_1 \\ 
         Y &\sim \mathcal{N}(1.2 + 0.25A + 0.5W_1 + AZ + 0.5Z + 0.4AW_2 - 0.75AW_3, 0)
         \end{aligned}$  \\
    \bottomrule
    \end{tabular}
    \label{tab:simdgm}
\end{table}
\addtolength{\tabcolsep}{-2pt}

DGM 1 considers a scenario where incorporating additional assumptions \ref{ass:effmod} and \ref{ass:popdiff} results in worse rather than better efficiency. In this case, using $\tilde\theta$ or $\tilde{\lambda}_C$ instead of $\tilde\lambda$ should yield a larger variance, which is the result of the positive correlation between the variance of the outcome conditional on $A$ and $W$ in the source population, $\sigma^2(A, W, S=1)$, and $\P(S=1\mid W)$ that is greater than the correlation between $\sigma^2(A, W, S=1)$ and $\P(S=1\mid Z)$.
DGM 2, is similar to DGM 1, but the correlation between $\sigma^2(A, W, S = 1)$ and $\P(S = 1 \mid W)$ is negative, resulting in $\tilde\theta$ and $\tilde\lambda_C$ having a smaller efficiency bound than $\tilde\lambda$.  DGM 3 is a special case where 
$V=Z$, and is formulated so that estimating $P(S = 1\mid W)$ may result in practical positivity violations while $P(S =1 \mid V)$ will not. DGM 4 satisfies both assumptions \ref{ass:effmod} and \ref{ass:popdiff}. We would expect estimators $\tilde{\theta}$ and $\tilde{\lambda}_C$ to also have smaller variance than $\tilde{\lambda}$ in DGMs 3 and 4.

We conducted 1000 simulations for sample sizes $n\in \{100,1000,10000\}$. All estimator nuisance parameters were correctly specified using an ensemble of a main-effects generalized linear model (GLM), a GLM including all two-way interactions, and an intercept-only model; random forests \citep{breiman2001random} and multivariate adaptive regression splines (MARS) \citep{friedman1991multivariate} were also included in the ensemble library when estimating $\tilde\lambda_C$. Estimator performance was evaluated in terms of absolute bias, 95\% confidence interval (CI) coverage, the variance between simulations scaled by $n$, the mean of the estimated variance within simulations, and the relative efficiency of the estimated variance comparing $\tilde \theta$ and $\tilde\lambda_C$ to $\tilde \lambda$. 

Table \ref{tab:sim_results_1_2} shows the results of our simulation using the one-step estimators, $\tilde \lambda, \tilde \theta, \tilde \lambda_C$, 
for DGMs 1 and 2. Table \ref{tab:sim_results_3_4} shows the results of our simulation using the one-step estimators, $\tilde \lambda, \tilde \theta, \tilde \lambda_C$, 
for DGMs 3 and 4. As expected, all estimators are unbiased or become unbiased with increasing sample size. With DGM 1, we see that estimating $\theta$ would incur larger variance as compared to estimating $\lambda$ due to the positive correlation between $\sigma^2(A, W, S = 1)$ and $\P(S = 1|W)$ that is greater than the correlation between $\sigma^2(A, W, S = 1)$ and $\P(S = 1|Z)$. However, when the correlation between $\sigma^2(A, W, S = 1)$ and $\P(S = 1|W)$ is switched to be negative in DGM 2 estimating $\theta$ instead of $\lambda$ yields efficiency gains. When $n=100$ all estimators are biased for with DGM 3; however, using $\tilde \theta$ results in improved confidence interval coverage and efficiency compared to $\tilde \lambda$, due to the practical positivity violations that are more of an issue when using $\tilde \lambda$. When the sample size is increased to $1000$ the efficiency gains of $\tilde\theta$ relative to $\tilde \lambda$ are more pronounced. When $V$ is treated as unknown, using $\tilde \lambda_C$ still yields efficiency gains compared to assuming all of $W$ are effect-modifiers.
DGM 4 represents a more complex, nonlinear outcome model. This may account for the suboptimal performance of $\tilde \theta$ and $\tilde \lambda_C$ compared to $\tilde \lambda$ under sample size $100$. When increasing the sample size to $1000$, the advantage of $\tilde \theta$ vs.~$\tilde \lambda$ becomes apparent. Finally, under the largest sample size of $10000$, both $\tilde \theta$ and $\tilde \lambda_C$ demonstrate marked efficiency gains over $\tilde \lambda$. 
We provide results without cross-fitting in Section S4 of the Supplement.

\begin{table}[H]
\centering
\caption{Simulation results comparing various ``one-step'' estimators at increasing sample-sizes among two data-generating mechanisms. There is a positive correlation between $\sigma^2(A, W, S = 1)$ and $\P(S = 1|W)$ in DGM 1, and a negative correlation between $\sigma^2(A, W, S = 1)$ and $\P(S = 1|W)$ in DGM 2.}
\label{tab:sim_results_1_2}
\footnotesize
\begin{tabular}[t]{ccccccc}
\toprule
Estimator & $n$ & $|\text{Bias}|$ & 95\% CI Covr. & $n \times$ Var. & Var. & Rel. Eff.\\
\midrule
\multicolumn{7}{l}{\textbf{DGM 1}} \\
$\tilde \lambda$ &  & 0.05 & 0.90 & 50.62 & 53.72 & 1.00\\
$\tilde \theta$ &  & 0.03 & 0.94 & 18.61 & 22.84 & 0.43\\
$\tilde \lambda_C$ & \multirow{-3}{*}{\centering\arraybackslash 100} & 0.01 & 0.91 & 37.64 & 41.14 & 0.77\\
\rule{0pt}{4ex}
$\tilde \lambda$ &  & 0.00 & 0.95 & 5.48 & 5.62 & 1.00\\
$\tilde \theta$ &  & 0.00 & 0.94 & 11.30 & 10.89 & 1.94\\
$\tilde \lambda_C$ & \multirow{-3}{*}{\centering\arraybackslash 1000} & 0.00 & 0.94 & 11.39 & 10.44 & 1.86\\
\rule{0pt}{4ex}
$\tilde \lambda$ &  & 0.00 & 0.96 & 4.82 & 5.11 & 1.00\\
$\tilde \theta$ &  & 0.00 & 0.95 & 9.61 & 10.65 & 2.08\\
$\tilde \lambda_C$ & \multirow{-3}{*}{\centering\arraybackslash 10000} & 0.00 & 0.95 & 9.55 & 10.63 & 2.08\\
\midrule
\multicolumn{7}{l}{\textbf{DGM 2}} \\
$\tilde \lambda$ &  & 0.02 & 0.89 & 47.02 & 40.43 & 1.00\\
$\tilde \theta$ &  & 0.01 & 0.94 & 4.81 & 4.67 & 0.12\\
$\tilde \lambda_C$ & \multirow{-3}{*}{\centering\arraybackslash 100} & 0.03 & 0.87 & 41.93 & 37.15 & 0.92\\
\rule{0pt}{4ex}
$\tilde \lambda$ &  & 0.00 & 0.95 & 13.17 & 13.30 & 1.00\\
$\tilde \theta$ &  & 0.00 & 0.96 & 2.82 & 3.05 & 0.23\\
$\tilde \lambda_C$ & \multirow{-3}{*}{\centering\arraybackslash 1000} & 0.00 & 0.96 & 5.61 & 4.77 & 0.36\\
\rule{0pt}{4ex}
$\tilde \lambda$ &  & 0.00 & 0.94 & 14.10 & 13.10 & 1.00\\
$\tilde \theta$ &  & 0.00 & 0.94 & 3.11 & 2.98 & 0.23\\
$\tilde \lambda_C$ & \multirow{-3}{*}{\centering\arraybackslash 10000} & 0.00 & 0.95 & 3.06 & 2.99 & 0.23\\
\bottomrule
\end{tabular}
\end{table}

\begin{table}[H]
\centering
\caption{Simulation results comparing ``one-step'' estimators, $\tilde \theta$, $\tilde \lambda$, and $\tilde \lambda_C$ at various sample-sizes with data-generating mechanisms 3 and 4.}
\label{tab:sim_results_3_4}
\footnotesize
\begin{tabular}[t]{ccccccc}
\toprule
Estimator & $n$ & $|\text{Bias}|$ & 95\% CI Covr. & $n \times$ Var. & Var. & Rel. Eff.\\
\midrule
\multicolumn{7}{l}{\textbf{DGM 3}} \\
$\tilde \lambda$ &  & 0.38 & 0.91 & 97.32 & 92.98 & 1.00\\
$\tilde \theta$ &  & 0.43 & 0.94 & 60.89 & 86.67 & 0.93\\
$\tilde \lambda_C$ & \multirow{-3}{*}{\centering\arraybackslash 100} & 0.10 & 0.90 & 106.41 & 92.40 & 0.99\\
\rule{0pt}{4ex}
$\tilde \lambda$ &  & 0.04 & 0.94 & 328.11 & 342.26 & 1.00\\
$\tilde \theta$ &  & 0.01 & 0.96 & 31.63 & 34.76 & 0.10\\
$\tilde \lambda_C$ & \multirow{-3}{*}{\centering\arraybackslash 1000} & 0.04 & 0.94 & 199.28 & 215.98 & 0.63\\
\rule{0pt}{4ex}
$\tilde \lambda$ &  & 0.00 & 0.94 & 110.27 & 105.26 & 1.00\\
$\tilde \theta$ &  & 0.00 & 0.95 & 33.98 & 32.06 & 0.30\\
$\tilde \lambda_C$ & \multirow{-3}{*}{\centering\arraybackslash 10000} & 0.00 & 0.94 & 48.54 & 45.38 & 0.43\\
\midrule
\multicolumn{7}{l}{\textbf{DGM 4}} \\
$\tilde \lambda$ &  & 0.20 & 0.95 & 49.45 & 52.68 & 1.00\\
$\tilde \theta$ &  & 0.36 & 0.93 & 116.34 & 101.57 & 1.93\\
$\tilde \lambda_C$ & \multirow{-4}{*}{\centering\arraybackslash 100} & 0.08 & 0.92 & 88.86 & 87.09 & 1.65\\
\rule{0pt}{4ex}
$\tilde \lambda$ &  & 0.01 & 0.91 & 41.00 & 34.58 & 1.00\\
$\tilde \theta$ &  & 0.00 & 0.94 & 26.87 & 27.14 & 0.78\\
$\tilde \lambda_C$ & \multirow{-4}{*}{\centering\arraybackslash 1000} & 0.01 & 0.91 & 67.71 & 59.30 & 1.71\\
\rule{0pt}{4ex}
$\tilde \lambda$ &  & 0.00 & 0.94 & 43.36 & 40.94 & 1.00\\
$\tilde \theta$ &  & 0.00 & 0.96 & 21.67 & 22.77 & 0.56\\
$\tilde \lambda_C$ & \multirow{-4}{*}{\centering\arraybackslash 10000} & 0.00 & 0.95 & 30.62 & 31.86 & 0.78\\
\bottomrule
\end{tabular}
\end{table}

\vspace{-.5cm}
\section{Motivating Example}
We now apply our proposed estimators to the Moving to Opportunity study (MTO), which was a randomized trial where families living in public housing in five U.S. cities could sign up to be randomized to receive a Section 8 voucher. 
These vouchers subsidize rent on the private market for families to be 30-40\% of their income. 
The families (parents and children) were followed for 10-15 years after randomized receipt (or not) of the housing voucher and educational, economic, and health outcomes were measured \citep{sanbonmatsu2011moving}. Here, we are interested in transporting the ATE of moving with the voucher ($A$) on behavioral problems in adolescence ($Y$, scored by the Behavioral Problems Index \citep{zill1990behavior}) from the New York City site ($S=1$, n=1200 (rounded sample size)) to the Los Angeles site ($S=0$, n=850 (rounded sample size)) among children who were $\le 5$ years old at the time of randomization. The two site-specific ATEs are qualitatively different (-0.0247 vs. 0.0012 for $S=1$ vs. $S=0$, respectively). It is possible that differences in the distribution of a sufficient subset of effect modifiers required for transport between the two cities, $Z$, can account for some of the difference in site-specific ATEs. We consider baseline covariates, $W$, that include parent and child characteristics and characteristics of the baseline neighborhood (a detailed list is in Section S5 of the Supplement). For the estimator $\tilde{\theta}$ that relies on specifying $V$ and $Z$, we include: $V=$\{gender, Black race, parent was under age 18 at birth of child, household member with a disability, parent graduated high school\}, and $Z=$\{household member with a disability, parent graduated high school\}. For this illustrative example, we use a single imputed dataset (missingness was $\le$2\% for covariates and 13\% for the outcome).  

\begin{figure}[hp]
\caption{Effect estimates and 95\% confidence intervals using data from the Moving to Opportunity Long-term Follow-up. The ATE is interpreted as the effect of moving with the housing voucher on Behavioral Problems Index score among adolescents, 10-15 years after randomization. New York City is considered as the source population and Los Angeles is considered as the target population in this example. ``Not transported'' denotes estimates using the non-transported one-step estimator; ``transportfullw'' denotes estimates using the transport one-step estimator $\tilde \lambda$; ``transport knownvz'' denotes estimates using our proposed transport one-step estimator assuming known effect modifiers and a known sufficient subset required for transport $\tilde \theta$; and ``transport unknownvz'' denotes estimates using our proposed transport collaborative one-step estimator when effect modifiers and a sufficient subset required for transport are unknown. All results were approved for release by the U.S. Census Bureau, authorization number CBDRB-FY23-CES018-009.}
\label{mtoresfig}
\centering\includegraphics[width=\textwidth,height=0.5\textheight,keepaspectratio]{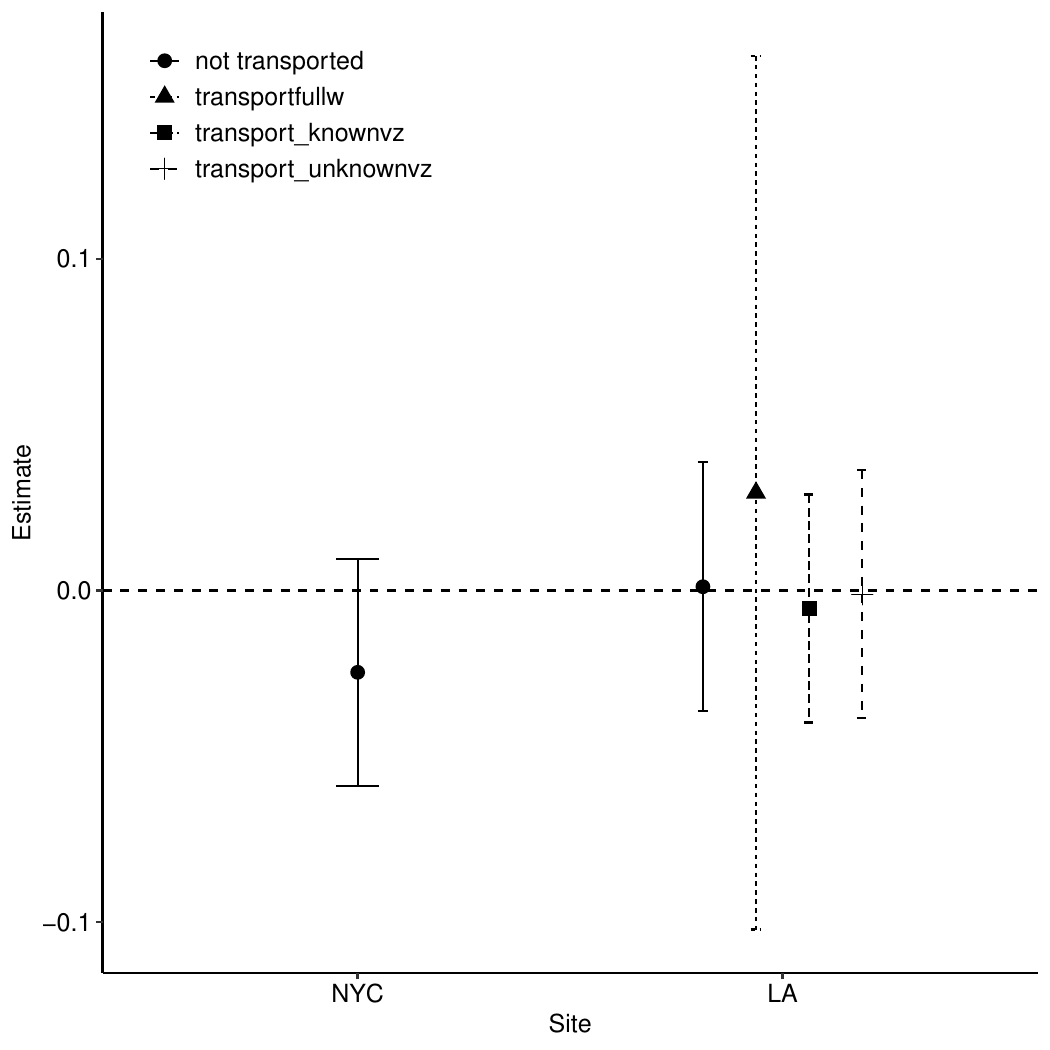}
\end{figure}

We see in Figure \ref{mtoresfig} that in New York City, moving with the housing voucher is associated with fewer behavioral problems among adolescent children, 10-15 years after baseline (risk difference (RD) -0.0245, 95\% CI: -0.0589, 0.0095). In contrast, in Los Angeles, moving with the housing voucher is not associated with fewer behavioral problems among adolescent children, 10-15 years after baseline (RD 0.0012, 95\% CI: -0.0364, 0.0387). To apply our transport estimators, we treat the NYC site as the source population and the LA site as the target population. This means that at this point, we do not use outcome data from LA. 

All three transport estimators correctly predict the null association in LA, but with appreciable differences in terms of their estimated variances. Using the existing one-step transport estimator that assumes all $W$ operates as effect modifiers and differ in distribution across sites, $\tilde \lambda$, we estimate a slightly positive effect in terms of the point estimate, but one that is decidedly non-statistically significant when taking the very wide confidence into account (RD: 0.0295, 95\% CI: -0.1022, 0.1611). In contrast, the confidence intervals using our proposed estimators are much narrower. Our one-step transport estimator that makes use of known $V$ and $Z$, $\tilde \theta$, estimates a null association with a confidence interval similar to the nontransported estimate (RD: -0.0054, 95\%CI: -0.0398, 0.0290). Similarly, even if we do not specify $V$ or $Z$, our collaborative one-step transport estimator, $\tilde \lambda_C$, estimate returns a very similar estimate with negligibly wider confidence intervals (RD: -0.0011, 95\% CI: -0.0385, 0.0363). 

\vspace{-.5cm}
\section{Conclusion}
We proposed semiparametric estimators to improve the precision of transported causal effects by considering two types of covariate dimension reduction. First, we proposed a collaborative estimator that that is agnostic as to the causal structure underlying effect heterogeneity but may offer efficiency gains by harnessing two insights: 1) for transport, it is sufficient to standardize the CATE with respect to the propensity score instead of with respect the full covariate vector, $W$; and 2) for transport, it is sufficient to re-weight the CATE by a propensity score that conditions only on the CATE. Second, we proposed two semiparametric estimators that incorporate knowledge of the causal structure underlying effect heterogeneity for transport. 
Our novel estimators demonstrated efficiency gains in many of the simulation scenarios considered as well as marked efficiency gains in the motivating example. This evidence, coupled with comparisons of the semiparametric efficiency bounds, suggests that our proposed estimators may offer meaningful efficiency gains over the existing semiparametric estimator \citep{rudolph2017robust} in real-world finite data analyses, particularly in observational studies challenged by practical positivity violations. Efficiency gains are not just a theoretical advantage---they can result in substantially narrower standard errors such that estimates convey more information and may make the difference between detecting versus not detecting a meaningful treatment effect. Moreover, because transport estimators may be used to predict the effect of a treatment or intervention if applied to a new, target population, their resulting estimates may be used to make policy decisions and inform population treatment guidelines \citep[e.g.,][]{matthay2022causal,mehrotra2021transporting}.  
Therefore, the ability to detect an effect may mean the difference between deciding to implement versus not implement policies that would improve population health and well-being.

\noindent \textbf{Acknowledgements:} 
This work was supported by R01DA056407.

This research was conducted as a part of the U.S. Census Bureau's Evidence Building Project Series. The Census Bureau has reviewed this data product to ensure appropriate access, use, and disclosure avoidance protection of the confidential source data used to produce this product (Data Management System (DMS) number:  P-7504667, Disclosure Review Board (DRB) approval number:  CBDRB-FY23-CES018-009).

\bibliographystyle{biom}
\vspace{-.5cm}
\bibliography{refs}

\begin{thebibliography}{}

\bibitem[\protect\citeauthoryear{Benkeser, Cai, and van~der Laan}{Benkeser
  et~al.}{2020}]{benkeser2020nonparametric}
Benkeser, D., Cai, W., and van~der Laan, M.~J. (2020).
\newblock A nonparametric super-efficient estimator of the average treatment
  effect.
\newblock {\em Statistical Science} {\bf 35,} 484--495.

\bibitem[\protect\citeauthoryear{Bennett, Vielma, and Zubizarreta}{Bennett
  et~al.}{2020}]{bennett2020building}
Bennett, M., Vielma, J.~P., and Zubizarreta, J.~R. (2020).
\newblock Building representative matched samples with multi-valued treatments
  in large observational studies.
\newblock {\em Journal of Computational and Graphical Statistics} {\bf 29,}
  744--757.

\bibitem[\protect\citeauthoryear{Breiman}{Breiman}{2001}]{breiman2001random}
Breiman, L. (2001).
\newblock Random forests.
\newblock {\em Machine learning} {\bf 45,} 5--32.

\bibitem[\protect\citeauthoryear{Chernozhukov, Chetverikov, Demirer, Duflo,
  Hansen, et~al\mbox{.}}{Chernozhukov et~al.}{2016}]{chernozhukov2016double}
Chernozhukov, V., Chetverikov, D., Demirer, M., Duflo, E., Hansen, C., et~al.
  (2016).
\newblock Double machine learning for treatment and causal parameters.
\newblock {\em arXiv preprint arXiv:1608.00060} .

\bibitem[\protect\citeauthoryear{Chernozhukov, Chetverikov, Demirer, Duflo,
  Hansen, Newey, and Robins}{Chernozhukov
  et~al.}{2018}]{chernozhukov2018double}
Chernozhukov, V., Chetverikov, D., Demirer, M., Duflo, E., Hansen, C., Newey,
  W., and Robins, J. (2018).
\newblock Double/debiased machine learning for treatment and structural
  parameters.
\newblock {\em The Econometrics Journal} {\bf 21,} C1--C68.

\bibitem[\protect\citeauthoryear{Cole and Stuart}{Cole and
  Stuart}{2010}]{cole2010generalizing}
Cole, S.~R. and Stuart, E.~A. (2010).
\newblock Generalizing evidence from randomized clinical trials to target
  populations the actg 320 trial.
\newblock {\em American journal of epidemiology} {\bf 172,} 107--115.

\bibitem[\protect\citeauthoryear{Colnet, Josse, Varoquaux, and Scornet}{Colnet
  et~al.}{2022}]{colnet2021generalizing}
Colnet, B., Josse, J., Varoquaux, G., and Scornet, E. (2022).
\newblock Causal effect on a target population: a sensitivity analysis to
  handle missing covariates.
\newblock {\em Journal of Causal Inference} {\bf 10,} 372--414.

\bibitem[\protect\citeauthoryear{Dahabreh, Robertson, Tchetgen, Stuart, and
  Hern{\'a}n}{Dahabreh et~al.}{2019}]{dahabreh2019generalizing}
Dahabreh, I.~J., Robertson, S.~E., Tchetgen, E.~J., Stuart, E.~A., and
  Hern{\'a}n, M.~A. (2019).
\newblock Generalizing causal inferences from individuals in randomized trials
  to all trial-eligible individuals.
\newblock {\em Biometrics} {\bf 75,} 685--694.

\bibitem[\protect\citeauthoryear{Egami and Hartman}{Egami and
  Hartman}{2021}]{egami2021covariate}
Egami, N. and Hartman, E. (2021).
\newblock Covariate selection for generalizing experimental results:
  application to a large-scale development program in uganda.
\newblock {\em Journal of the Royal Statistical Society Series A: Statistics in
  Society} {\bf 184,} 1524--1548.

\bibitem[\protect\citeauthoryear{Friedman et~al\mbox{.}}{Friedman
  et~al.}{1991}]{friedman1991multivariate}
Friedman, J.~H. et~al. (1991).
\newblock Multivariate adaptive regression splines.
\newblock {\em The annals of statistics} {\bf 19,} 1--67.

\bibitem[\protect\citeauthoryear{Hahn, Murray, and Carvalho}{Hahn
  et~al.}{2020}]{hahn2020bayesian}
Hahn, P.~R., Murray, J.~S., and Carvalho, C.~M. (2020).
\newblock Bayesian regression tree models for causal inference: Regularization,
  confounding, and heterogeneous effects (with discussion).
\newblock {\em Bayesian Analysis} {\bf 15,} 965--1056.

\bibitem[\protect\citeauthoryear{Hern{\'a}n, Hern{\'a}ndez-D{\'\i}az, Werler,
  and Mitchell}{Hern{\'a}n et~al.}{2002}]{hernan2002causal}
Hern{\'a}n, M.~A., Hern{\'a}ndez-D{\'\i}az, S., Werler, M.~M., and Mitchell,
  A.~A. (2002).
\newblock Causal knowledge as a prerequisite for confounding evaluation: an
  application to birth defects epidemiology.
\newblock {\em American journal of epidemiology} {\bf 155,} 176--184.

\bibitem[\protect\citeauthoryear{Kennedy}{Kennedy}{2022}]{kennedy2022semiparametric}
Kennedy, E.~H. (2022).
\newblock Semiparametric doubly robust targeted double machine learning: a
  review.
\newblock {\em arXiv preprint arXiv:2203.06469} .

\bibitem[\protect\citeauthoryear{Klaassen}{Klaassen}{1987}]{klaassen1987consistent}
Klaassen, C.~A. (1987).
\newblock Consistent estimation of the influence function of locally
  asymptotically linear estimators.
\newblock {\em The Annals of Statistics} pages 1548--1562.

\bibitem[\protect\citeauthoryear{Li, Morgan, and Zaslavsky}{Li
  et~al.}{2018}]{li2018balancing}
Li, F., Morgan, K.~L., and Zaslavsky, A.~M. (2018).
\newblock Balancing covariates via propensity score weighting.
\newblock {\em Journal of the American Statistical Association} {\bf 113,}
  390--400.

\bibitem[\protect\citeauthoryear{Luedtke and van~der Laan}{Luedtke and van~der
  Laan}{2016}]{luedtke2016super}
Luedtke, A.~R. and van~der Laan, M.~J. (2016).
\newblock Super-learning of an optimal dynamic treatment rule.
\newblock {\em The international journal of biostatistics} {\bf 12,} 305--332.

\bibitem[\protect\citeauthoryear{Matthay and Glymour}{Matthay and
  Glymour}{2022}]{matthay2022causal}
Matthay, E.~C. and Glymour, M.~M. (2022).
\newblock Causal inference challenges and new directions for epidemiologic
  research on the health effects of social policies.
\newblock {\em Current Epidemiology Reports} {\bf 9,} 22--37.

\bibitem[\protect\citeauthoryear{Mehrotra, Westreich, Glymour, Geng, and
  Glidden}{Mehrotra et~al.}{2021}]{mehrotra2021transporting}
Mehrotra, M.~L., Westreich, D., Glymour, M.~M., Geng, E., and Glidden, D.~V.
  (2021).
\newblock Transporting subgroup analyses of randomized controlled trials for
  planning implementation of new interventions.
\newblock {\em American journal of epidemiology} {\bf 190,} 1671--1680.

\bibitem[\protect\citeauthoryear{Nie and Wager}{Nie and
  Wager}{2021}]{nie2021quasi}
Nie, X. and Wager, S. (2021).
\newblock Quasi-oracle estimation of heterogeneous treatment effects.
\newblock {\em Biometrika} {\bf 108,} 299--319.

\bibitem[\protect\citeauthoryear{Pearl}{Pearl}{2009}]{Pearl2009}
Pearl, J. (2009).
\newblock {Myth, Confusion, and Science in Causal Analysis}.
\newblock Technical Report R-348, Cognitive Systems Laboratory, Computer
  Science Department University of California, Los Angeles, Los Angeles, CA.

\bibitem[\protect\citeauthoryear{Pearl}{Pearl}{2015}]{pearl2015generalizing}
Pearl, J. (2015).
\newblock Generalizing experimental findings.
\newblock {\em Journal of Causal Inference} {\bf 3,} 259--266.

\bibitem[\protect\citeauthoryear{Pearl and Bareinboim}{Pearl and
  Bareinboim}{2011}]{pearl2011transportability}
Pearl, J. and Bareinboim, E. (2011).
\newblock Transportability of causal and statistical relations: A formal
  approach.
\newblock In {\em Twenty-fifth AAAI conference on artificial intelligence}.

\bibitem[\protect\citeauthoryear{Rosenbaum and Rubin}{Rosenbaum and
  Rubin}{1983}]{rosenbaum1983central}
Rosenbaum, P.~R. and Rubin, D.~B. (1983).
\newblock The central role of the propensity score in observational studies for
  causal effects.
\newblock {\em Biometrika} {\bf 70,} 41--55.

\bibitem[\protect\citeauthoryear{Rubin and van~der Laan}{Rubin and van~der
  Laan}{2007}]{rubin2007doubly}
Rubin, D. and van~der Laan, M.~J. (2007).
\newblock A doubly robust censoring unbiased transformation.
\newblock {\em The international journal of biostatistics} {\bf 3,}.

\bibitem[\protect\citeauthoryear{Rudolph and van~der Laan}{Rudolph and van~der
  Laan}{2017}]{rudolph2017robust}
Rudolph, K.~E. and van~der Laan, M.~J. (2017).
\newblock Robust estimation of encouragement-design intervention effects
  transported across sites.
\newblock {\em Journal of the Royal Statistical Society. Series B, Statistical
  methodology} {\bf 79,} 1509.

\bibitem[\protect\citeauthoryear{Sanbonmatsu, Katz, Ludwig, Gennetian, Duncan,
  Kessler, Adam, McDade, and Lindau}{Sanbonmatsu
  et~al.}{2011}]{sanbonmatsu2011moving}
Sanbonmatsu, L., Katz, L.~F., Ludwig, J., Gennetian, L.~A., Duncan, G.~J.,
  Kessler, R.~C., Adam, E.~K., McDade, T., and Lindau, S.~T. (2011).
\newblock Moving to opportunity for fair housing demonstration program: Final
  impacts evaluation.

\bibitem[\protect\citeauthoryear{Schnitzer and Cefalu}{Schnitzer and
  Cefalu}{2018}]{schnitzer2018collaborative}
Schnitzer, M.~E. and Cefalu, M. (2018).
\newblock Collaborative targeted learning using regression shrinkage.
\newblock {\em Statistics in Medicine} {\bf 37,} 530--543.

\bibitem[\protect\citeauthoryear{Stuart, Cole, Bradshaw, and Leaf}{Stuart
  et~al.}{2011}]{stuart2011use}
Stuart, E.~A., Cole, S.~R., Bradshaw, C.~P., and Leaf, P.~J. (2011).
\newblock The use of propensity scores to assess the generalizability of
  results from randomized trials.
\newblock {\em Journal of the Royal Statistical Society: Series A (Statistics
  in Society)} {\bf 174,} 369--386.

\bibitem[\protect\citeauthoryear{van~der Laan and Gruber}{van~der Laan and
  Gruber}{2010}]{van2010collaborative}
van~der Laan, M.~J. and Gruber, S. (2010).
\newblock Collaborative double robust targeted maximum likelihood estimation.
\newblock {\em The international journal of biostatistics} {\bf 6,}.

\bibitem[\protect\citeauthoryear{{von Mises}}{{von
  Mises}}{1947}]{mises1947asymptotic}
{von Mises}, R. (1947).
\newblock On the asymptotic distribution of differentiable statistical
  functions.
\newblock {\em The annals of mathematical statistics} {\bf 18,} 309--348.

\bibitem[\protect\citeauthoryear{Zeng, Kennedy, Bodnar, and Naimi}{Zeng
  et~al.}{2023}]{zeng2023efficient}
Zeng, Z., Kennedy, E.~H., Bodnar, L.~M., and Naimi, A.~I. (2023).
\newblock Efficient generalization and transportation.
\newblock {\em arXiv preprint arXiv:2302.00092} .

\bibitem[\protect\citeauthoryear{Zheng and van~der Laan}{Zheng and van~der
  Laan}{2011}]{zheng2011cross}
Zheng, W. and van~der Laan, M.~J. (2011).
\newblock Cross-validated targeted minimum-loss-based estimation.
\newblock In {\em Targeted Learning}, pages 459--474. Springer.

\bibitem[\protect\citeauthoryear{Zill}{Zill}{1990}]{zill1990behavior}
Zill, N. (1990).
\newblock {\em Behavior problems index based on parent report}.
\newblock Child Trends.

\end{thebibliography}


\newpage

\newtheorem{innercustomgeneric}{\customgenericname}
\providecommand{\customgenericname}{}
\newcommand{\newcustomtheorem}[2]{%
  \newenvironment{#1}[1]
  {%
   \renewcommand\customgenericname{#2}%
   \renewcommand\theinnercustomgeneric{##1}%
   \innercustomgeneric
  }
  {\endinnercustomgeneric}
}

\newcustomtheorem{customthm}{Theorem}
\newcustomtheorem{customlemma}{Lemma}
\renewcommand{\figurename}{Figure S}
\renewcommand{\tablename}{Table S}
\renewcommand{\thesection}{S\arabic{section}}
\setcounter{figure}{0} 
\setcounter{table}{0} 
\setcounter{section}{0} 
\section*{Supplementary Materials}

\section{Proof of identification result in Lemma 1.} 
\begin{proof}
\begin{align*}
    \E(Y_1-Y_0\mid S=0)&=\E\{\E(Y_1-Y_0\mid W, S=0)\mid S=0\}\\
    &\text{by assumption 1}\\
    &=\E\{\E(Y_1-Y_0\mid A, W, S=0)\mid S=0\}\\
        &\text{by consistency}\\
        &=\E\{\E(Y\mid A=1, W, S=0)-\E(Y\mid A=0, W, S=0)\mid S=0\}\\
         &\text{by assumption 2}\\
                &=\E\{\E(Y\mid A=1, W, S=1)-\E(Y\mid A=0, W, S=1)\mid S=0\}\\
                &=\E\{f(W)\mid S=0\}\\
                &\text{by assumption 3}\\
                                &=\E\{f(V)\mid S=0\}\\ 
        &\text{by assumption 4}\\
&=\E\{\E[f(V)\mid Z, S=0]\mid S=0\}\\ 
&=\E\{\E[f(V)\mid Z, S=1]\mid S=0\}\\
&=\E\{\E[f(V)\mid Z]\mid S=0\}.
\end{align*}
\end{proof}

\section{Proof of Theorem 1}
    \begin{proof}\allowdisplaybreaks
    We will use the notation $k(W)=\E\{f(W)\mid e_S(W)\}$ in this proof.
{\small
\begin{align}
    \E[\U_\lambda(O;\hat\eta)] &=\E\left[\frac{\one\{S=1\}}{\p}\left\{\frac{A}{\hat e_A(W)}-\frac{1-A}{1-\hat e_A(W)}\right\}\frac{1-\hat h_S(W)}{\hat h_S(W)}\{Y -
                        A \hat f(W) - \hat g(W)\right]\notag\\
                        &+\E\left[\frac{1-\hat e_S(W)}{\p}\{\hat f(W)-\hat k(W)\}\right]\notag\\
                        &+\E\left[\frac{1-e_S(W)}{\p}\{\hat k(W)- k(W)\}\right]\notag\\
                        &=\E\left[\frac{\one\{S=1\}}{\p}\left\{\frac{A}{\hat e_A(W)}-\frac{1-A}{1-\hat e_A(W)}\right\}\frac{1-\hat h_S(W)}{\hat h_S(W)}\{A (f(W) - \hat f(W)) + g(W)- \hat g(W)\right]\notag\\
                        &+\E\left[\frac{1-e_S(W)}{\p}\{\hat f(W)-f(W)\}\right]\notag\\
                        &+\E\left[\frac{\hat e_S(W)-e_S(W)}{\p}\{\hat f(W)-\hat k(W)\}\right]\notag\\
                        &=\E\left[\frac{\one\{S=1\}}{\p}    \frac{A}{\hat e_A(W)}\frac{1-\hat h_S(W)}{\hat h_S(W)}\{f(W) - \hat f(W) + g(W)- \hat g(W)\}\right]\notag\\
                                                &-\E\left[\frac{\one\{S=1\}}{\p}    \frac{1-A}{1-\hat e_A(W)}\frac{1-\hat h_S(W)}{\hat h_S(W)}\{g(W)- \hat g(W)\}\right]\notag\\
                        &+\E\left[\frac{1-e_S(W)}{\p}\{\hat f(W)-f(W)\}\right]\notag\\
                          &+\E\left[\frac{\hat e_S(W)-e_S(W)}{\p}\{\E[\hat f(W)\mid \hat e_S(W),e_S(W)]-\hat k(W)\}\right]\label{eq:t1}\\                        &=\E\left[\frac{e_S(W)}{\p}    \frac{e_A(W)}{\hat e_A(W)}\frac{1-\hat h_S(W)}{\hat h_S(W)}\{f(W) - \hat f(W) + g(W)- \hat g(W)\}\right]\notag\\
                                                &-\E\left[\frac{e_S(W)}{\p}    \frac{1-e_A(W)}{1-\hat e_A(W)}\frac{1-\hat h_S(W)}{\hat h_S(W)}\{g(W)- \hat g(W)\}\right]\notag\\
            &+\E\left[\frac{1-e_S(W)}{\p}\{\hat f(W)-f(W)\}\right]+(\ref{eq:t1})\notag\\
                &=\E\left[\frac{e_S(W)}{\p}\frac{1-\hat h_S(W)}{\hat h_S(W)}    \left\{\frac{e_A(W)}{\hat e_A(W)}-\frac{1-e_A(W)}{1-\hat e_A(W)}\right\}\{g(W)- \hat g(W)\}\right]\label{eq:first}\\
                        &+\E\left[\left\{\frac{e_S(W)}{\p}    \frac{e_A(W)}{\hat e_A(W)}\frac{1-\hat h_S(W)}{\hat h_S(W)} - \frac{1-e_S(W)}{\p}\right\}\{f(W)-\hat f(W)\}\right]+(\ref{eq:t1})\notag\\
                                        &=(\ref{eq:first})\notag\\                                    &+\E\left[\frac{e_S(W)}{\p}\frac{1-\hat h_S(W)}{\hat h_S(W)}    \left\{\frac{e_A(W)}{\hat e_A(W)}-1\right\}\{f(W)- \hat f(W)\}\right]\label{eq:second}\\
                        &+\E\left[\left\{\frac{e_S(W)}{\p}\frac{1-\hat h_S(W)}{\hat h_S(W)} - \frac{1-e_S(W)}{\p}\right\}\{f(W)-\hat f(W)\}\right]+(\ref{eq:t1})\notag\\
                                        &=(\ref{eq:t1})+(\ref{eq:first})+(\ref{eq:second})
                        +\E\left[\left\{\frac{e_S(W)}{\p}\frac{1-\hat h_S(W)}{\hat h_S(W)} - \frac{1-e_S(W)}{\p}\right\}\{f(W)-\hat f(W)\}\right]\notag\\
                                        &=(\ref{eq:t1})+(\ref{eq:first})+(\ref{eq:second})+\E\left[\frac{1}{\p\times \hat h_S(W)}\left\{e_S(W)-\hat h_S(W)\right\}\{f(W)-\hat f(W)\}\right]\notag\\
                         &=(\ref{eq:t1})+(\ref{eq:first})+(\ref{eq:second})+\E\left[\frac{1}{\p\times \hat h_S(W)}\left\{S-\hat h_S(W)\right\}\{f(W)-\hat f(W)\}\right]\notag\\
                        &=(\ref{eq:t1})+(\ref{eq:first})+(\ref{eq:second})\notag\\
                        &+\E\left[\frac{1}{\p\times \hat h_S(W)}\left\{h^*_S(W)-\hat h_S(W)\right\}\{f(W)-\hat f(W)\}\right],\label{eq:third}
\end{align}
}%
where we define $h^*_S(w)=\P(S=1\mid f(W)=f(w), \hat f(W)=\hat f(w))$.
The theorem follows after noticing that 
\begin{align}
\E[\U_{\lambda}(O;\hat\eta) - \U_{\hat\lambda}(O;\hat\eta)]&=\hat\lambda-\lambda\notag\\
&+\left\{\frac{\p}{\hat\p}-1\right\}(\hat\lambda-\lambda)\label{eq:fourth}.
\end{align}
where we define $R_U(\hat\eta,\eta)=(\ref{eq:t1})+(\ref{eq:first})+(\ref{eq:second})+(\ref{eq:third})+(\ref{eq:fourth})$.
\end{proof}

\section{Estimator for $\theta_{\text{alt}}$}
Here, we propose an estimator for $\theta_{\text{alt}}$, defined below. This estimator only requires that $Z$ is measured in the target population, $S=0$, so it will be necessary to use $\theta_{\text{alt}}$ instead of $\theta$ when $V \backslash Z$ is not measured in the target population. 

\[\theta_{\text{alt}} = \E[\E\{f(V)\mid Z, S=1\}\mid S=0].\]

As the proposed estimator is based on the influence function, we first give the following result.
\begin{customthm}{4}
\label{lemma:eifalt}
 The function $\D_{\theta,\text{alt}}(O;\P)$, defined below, satisfies Definition 1 in the model with restrictions A3 and A4. 
  \begin{align*}
      \D_{\theta,\text{alt}}(O;\P) = \frac{1}{\P(S=0)}\bigg[&\frac{\one\{S=1\}(2A-1)}{\P(A\mid S=1,
                W)}\frac{\P(S=0\mid Z)}{\P(S=1\mid Z)}\{Y -
                        A f(V) - g(W)\}\\
&+\frac{\one\{S=1\}\P(S=0\mid Z)}{\P(S=1\mid Z)}[f(V)-\E\{f(V)\mid Z, S=1\}]\\
               &+\one\{S=0\}[\E\{f(V)\mid Z, S=1\}-\theta_{\text{alt}}]\bigg].
  \end{align*}
  \end{customthm}

We propose a one-step semiparametric estimator of $\theta_{\text{alt}}$ under Assumptions 3 and 4, constructed by subtracting the first-order bias estimate from the plug-in estimator: $\tilde \theta_{\text{alt}} = \hat \theta_{\text{alt}}(O, \hat\P) + \frac{1}{n}\sum^n_{i=1}\D_{\theta, \text{alt}}(O_i, \hat \gamma).$ We use a doubly robust unbiased transformation to estimate
$f(V)$, and another doubly robust unbiased transformation to estimate $\E\{f(V) \mid Z, S=1\}$, and then plug these estimates into the one-step estimator. This 
makes our estimator doubly robust. We use Lemma 1 from the main text (under assumption A3) and the following: 

\begin{customlemma}{3}[Doubly robust unbiased transformation for estimation of $\E\{f(V)\mid Z, S=1\}$] 
\label{lemma:dref}For any
  distribution $\P_1$, define
  \begin{align*}
    \B(O;\P_1) &= \frac{\one\{S=1\}}{\P_1(S=1\mid 
      Z)}\{f(V) - \E_1(f(V)\mid Z, S=1)\} + \E_1(f(V)\mid Z, S=1).
  \end{align*}
  
  If $\P_1$ is such that $\P_1(S=1\mid Z)=\P(S=1\mid Z)$ or $\E_1(f(V)\mid Z, S=1)=\E(f(V)\mid Z, S=1)$, we have that $\E\{\B(O;\P_1)\mid Z\}=\E\{f(W)\mid Z, S=1\}$.
\end{customlemma}

Our estimation strategy proceeds as follows: 
\begin{enumerate}
\item First, estimate nuisance parameters $\hat\P(A\mid S=1, W)$ and
  $\hat\E(Y\mid A, W, S=1)$, as described in the main text.  
\item Use the above nuisance parameter estimates to calculate $\T(O_i;\hat\P)$ for observations with $S_i=1$.
\item Regress $\T(O_i;\hat\P)$ on $V_i$ among observations with $S_i=1$. The resulting predicted values are estimates of $f(V_i)$, denoted $\hat f(V_i)$.
\item Regress $\hat f(V_i)$ on $Z_i$ in the source population. The resulting predicted values are denoted $\hat\E\{f(V_i)\mid Z_i, S=1\}$.
\item Compute $\B(O_i; \hat \P)$ for observations with $S_i=1$.
\item Regress $\B(O_i; \hat \P)$ on $Z_i$.  Using this fitted model, generate predicted values for all observations $i$, denoted $\hat\E\{\B(O_i; \hat \P)\mid Z_i\}$.
\item Compute an initial estimator $\hat \theta_{alt} = \frac{1}{\sum_{i=1}^n
    \one\{S_i=0\}}\sum_{i=1}^n
  \one\{S_i=0\}\hat
  \E\{\B(O_i; \hat \P)\mid Z_i\}$.
\item Compute estimates of $g(W_i)$ for observations with $S_i=1$, denoted $\hat g(W_i)$, by generating predicted values from the estimator of $\E(Y\mid A=0, W, S=1)$.  
\item Compute the one-step estimator as $\tilde\theta_{alt} = \hat \theta_{alt}
  +\frac{1}{n}\sum \D_{\theta, alt}(O_i;\hat \P)$, where $\hat P$ is comprised of
   $\hat P(A\mid S=1, W)$, $\hat P(S=1\mid Z)$,
  $\hat f(V)$, $\hat g(W)$ and $\hat\E\{\hat f(V)\mid Z\}$.
  \item Last, we can estimate the variance of $\tilde\theta_{alt}$ as the sample variance of $\D_{\tilde \theta, alt}(O_i;\hat \P)$.
\end{enumerate}

\section{Additional simulation results}

\begin{table}[H]
\centering
\caption{Simulation results comparing one-step estimators, $\tilde \theta$, $\tilde \lambda$, and $\tilde \lambda_C$ at various sample-sizes with data-generating mechanisms 1 and 2 with no cross-fitting.}
\label{tab:sim_results_3_4_nocf}
\footnotesize
\begin{tabular}[t]{ccccccc}
\toprule
Estimator & $n$ & $|\text{Bias}|$ & 95\% CI Covr. & $n \times$ Var. & Var. & Rel. Eff.\\
\midrule
\multicolumn{7}{l}{\textbf{DGM 1}} \\
$\tilde \lambda$ &  & 0.09 & 0.75 & 11.71 & 5.16 & 1.00\\
$\tilde \theta$ &  & 0.04 & 0.86 & 13.92 & 8.65 & 1.68\\
$\tilde \lambda_C$ & \multirow{-3}{*}{\centering\arraybackslash 100} & 0.05 & 0.84 & 13.01 & 8.03 & 1.55\\
\rule{0pt}{4ex}
$\tilde \lambda$ &  & 0.00 & 0.95 & 5.39 & 5.26 & 1.00\\
$\tilde \theta$ &  & 0.00 & 0.95 & 10.66 & 10.39 & 1.97\\
$\tilde \lambda_C$ & \multirow{-3}{*}{\centering\arraybackslash 1000} & 0.00 & 0.95 & 9.44 & 9.87 & 1.87\\
\rule{0pt}{4ex}
$\tilde \lambda$ &  & 0.00 & 0.95 & 5.13 & 5.09 & 1.00\\
$\tilde \theta$ &  & 0.00 & 0.95 & 10.48 & 10.63 & 2.09\\
$\tilde \lambda_C$ & \multirow{-3}{*}{\centering\arraybackslash 10000} & 0.00 & 0.95 & 10.48 & 10.61 & 2.08\\
\midrule
\multicolumn{7}{l}{\textbf{DGM 2}} \\
$\tilde \lambda$ &  & 0.02 & 0.67 & 15.08 & 4.09 & 1.00\\
$\tilde \theta$ &  & 0.01 & 0.91 & 3.66 & 2.72 & 0.67\\
$\tilde \lambda_C$ & \multirow{-3}{*}{\centering\arraybackslash 100} & 0.01 & 0.70 & 11.97 & 2.95 & 0.72\\
\rule{0pt}{4ex}
$\tilde \lambda$ &  & 0.00 & 0.93 & 12.65 & 11.11 & 1.00\\
$\tilde \theta$ &  & 0.00 & 0.96 & 2.91 & 2.93 & 0.26\\
$\tilde \lambda_C$ & \multirow{-3}{*}{\centering\arraybackslash 1000} & 0.00 & 0.93 & 6.95 & 4.12 & 0.37\\
\rule{0pt}{4ex}
$\tilde \lambda$ &  & 0.00 & 0.95 & 13.36 & 12.94 & 1.00\\
$\tilde \theta$ &  & 0.00 & 0.96 & 2.91 & 2.97 & 0.23\\
$\tilde \lambda_C$ & \multirow{-3}{*}{\centering\arraybackslash 10000} & 0.00 & 0.96 & 2.91 & 2.97 & 0.23\\
\bottomrule
\end{tabular}
\end{table}

\begin{table}[H]
\centering
\caption{Simulation results comparing ``one-step'' estimators, $\tilde \theta$, $\tilde \lambda$, and $\tilde \lambda_C$ at various sample-sizes with data-generating mechanisms 3 and 4 with no cross-fitting.}
\label{tab:sim_results_3_4_nocf}
\footnotesize
\begin{tabular}[t]{ccccccc}
\toprule
Estimator & $n$ & $|\text{Bias}|$ & 95\% CI Covr. & $n \times$ Var. & Var. & Rel. Eff.\\
\midrule
\multicolumn{7}{l}{\textbf{DGM 3}} \\
$\tilde \lambda$ &  & 0.00 & 0.82 & 26.81 & 18.07 & 1.00\\
$\tilde \theta$ &  & 0.12 & 0.86 & 32.18 & 26.10 & 1.44\\
$\tilde \lambda_C$ & \multirow{-3}{*}{\centering\arraybackslash 100} & 0.01 & 0.75 & 35.76 & 15.87 & 0.88\\
\rule{0pt}{4ex}
$\tilde \lambda$ &  & 0.02 & 0.86 & 45.85 & 32.42 & 1.00\\
$\tilde \theta$ &  & 0.00 & 0.94 & 31.71 & 30.87 & 0.95\\
$\tilde \lambda_C$ & \multirow{-3}{*}{\centering\arraybackslash 1000} & 0.00 & 0.91 & 38.69 & 32.30 & 1.00\\
\rule{0pt}{4ex}
$\tilde \lambda$ &  & 0.00 & 0.95 & 94.12 & 90.24 & 1.00\\
$\tilde \theta$ &  & 0.00 & 0.95 & 32.01 & 31.74 & 0.35\\
$\tilde \lambda_C$ & \multirow{-3}{*}{\centering\arraybackslash 10000} & 0.00 & 0.95 & 41.59 & 40.82 & 0.45\\
\midrule
\multicolumn{7}{l}{\textbf{DGM 4}} \\
$\tilde \lambda$ &  & 0.19 & 0.76 & 16.15 & 9.70 & 1.00\\
$\tilde \theta$ &  & 0.10 & 0.81 & 25.34 & 14.77 & 1.52\\
$\tilde \lambda_C$ & \multirow{-4}{*}{\centering\arraybackslash 100} & 0.13 & 0.77 & 26.42 & 13.41 & 1.38\\
\rule{0pt}{4ex}
$\tilde \lambda$ &  & 0.03 & 0.86 & 29.24 & 19.78 & 1.00\\
$\tilde \theta$ &  & 0.00 & 0.93 & 23.35 & 19.96 & 1.01\\
$\tilde \lambda_C$ & \multirow{-4}{*}{\centering\arraybackslash 1000} & 0.00 & 0.86 & 32.79 & 20.36 & 1.03\\
\rule{0pt}{4ex}
$\tilde \lambda$ &  & 0.00 & 0.94 & 42.31 & 39.61 & 1.00\\
$\tilde \theta$ &  & 0.00 & 0.95 & 22.12 & 22.06 & 0.56\\
$\tilde \lambda_C$ & \multirow{-4}{*}{\centering\arraybackslash 10000} & 0.00 & 0.94 & 29.89 & 27.35 & 0.69\\
\bottomrule
\end{tabular}
\end{table}

\section{MTO baseline covariates} We considered the following baseline covariates, $W$, that include parent and child characteristics and characteristics of the baseline neighborhood (with percent missing given in parentheses):
 \begin{itemize}
     \item Adolescent characteristics (all had 0\% missing except race/ethnicity, which had 2\% missing): site (LA, NYC), age, race/ethnicity (categorized as black, latine, white, other), number of family members 
(categorized as 2, 3, or 4+), someone from school asked to discuss problems the child had with schoolwork or behavior during the 2 years prior to baseline, child enrolled in special class for gifted and talented students.
     \item Adult household head characteristics (which all had 0\% missing): high school graduate, marital status (never vs ever married), whether had been a teen parent, work status, receipt of AFDC/TANF, whether any family member has a disability.
     \item Neighborhood characteristics (which all had 0\% missing except neighborhood poverty, which had 2\% missing): felt neighborhood streets were unsafe at night; very dissatisfied with neighborhood; poverty level of neighborhood.
     \item Reported reasons for participating in MTO (which had 0\% missing): to have access to better schools.
     \item Moving-related characteristics (which had 0\% missing): moved more then 3 times during the 5 years prior to baseline, previous application for Section 8 voucher.
 \end{itemize}

\end{document}